\documentclass[12pt,letterpaper]{article}
\usepackage[dvipsnames]{xcolor}
\usepackage{jheppub}
\usepackage[utf8]{inputenc}
\usepackage{amsthm}
\usepackage{dcolumn}
\usepackage{cancel}
\usepackage{booktabs}
\usepackage{multirow}
\usepackage{esvect}
\usepackage{braket}
\usepackage{float}
\usepackage{ragged2e}

\usepackage{tikz}
\usetikzlibrary{shapes,arrows,calc}
\usepackage{subcaption}

\usepackage{url}
\usepackage{hyperref}

\newcolumntype{C}[1]{>{\centering\let\newline\\\arraybackslash\hspace{0pt}}m{#1}}
\newtheorem{theorem}{Theorem}

\def\be{\begin{equation}}
\def\ee{\end{equation}}
\def\ba#1\ea{\begin{align}#1\end{align}}
\def\bg#1\eg{\begin{gather}#1\end{gather}}
\def\bm#1\em{\begin{multline}#1\end{multline}}
\def\bmd#1\emd{\begin{multlined}#1\end{multlined}}

\def\se{\section}

\def\({\left(}
\def\){\right)}
\def\[{\left[}
\def\]{\right]}
\def\<{\langle}
\def\>{\rangle}

\def\cA{{\mathcal A}}

\begin{document}
%Full Eigenstate Thermalization in Quantum Field Theory: Bounds and Gravitational Saturation
\title{Full Eigenstate Thermalization in Quantum Field Theory: Free Cumulants and Gravitational Scrambling}
\author[a]{Ricardo Espíndola,}
\emailAdd{ricardo.esro1@gmail.com}
\affiliation[a]{Institute for Advanced Study, Tsinghua University, Beijing 100084, China}
\author[b]{and Viktor Jahnke}
\emailAdd{v.jahnke@unesp.br}
\affiliation[b]{Instituto de Física Teórica, UNESP-Universidade Estadual Paulista
R. Dr. Bento T. Ferraz 271, Bl. II, Sao Paulo 01140-070, SP, Brazil}

\abstract{
Using thermal analyticity, we derive directional large-frequency bounds on the smooth functions entering full eigenstate thermalization in continuum quantum field theory. The strongest directions reproduce the scales previously identified for lattice systems, while the complete analyticity domain reveals a broader asymmetric multifrequency structure. In thermal two-dimensional conformal field theory, exact two- and three-point functions saturate the corresponding bounds. At fourth order, where genuine dynamical information first appears, we show that the holographic CFTs studied here saturate the strongest full-ETH bounds. These results establish a direct connection between full ETH, free cumulants, and gravitational scrambling, and are consistent with the emergence of asymptotic freeness realized by the gravitational scrambling algebra.

}

\maketitle

%%%%%%%%%%%%%%%%%%%%%%%%%%%%%%%%%%%%%%%%%%%%%%%%%%
\section{Introduction }

Complicated non-integrable dynamics in quantum systems often leads to universal statistical features that are usually referred to as ``quantum chaos.'' A paradigmatic example is random-matrix universality, namely the observation that quantum systems with classically chaotic counterparts display energy-level correlations similar to those of random matrix theory \cite{BGS}. Another important diagnostic is provided by out-of-time-order correlators (OTOCs), which capture the quantum analogue of the exponential sensitivity to initial conditions characteristic of classical chaotic dynamics \cite{Garcia-Mata:2022voo}. Somewhat surprisingly, these ideas have found important applications in high-energy physics, particularly in the context of the AdS/CFT correspondence. Seminal works showed that holographic systems display maximal scrambling \cite{Shenker:2013pqa,Shenker:2013yza,Shenker:2014cwa,Roberts:2014isa,Roberts:2014ifa} and that the Lyapunov exponent governing the growth of thermal OTOCs is bounded from above by $\lambda_L\leq 2\pi/\beta$, with holographic systems saturating this bound \cite{Maldacena:2015waa}.\footnote{See Ref.~\cite{Jahnke:2018off} for a pedagogical review.} More recently, quantum-chaotic universality has also become an important guide in the study of the quantum gravity path integral \cite{Altland:2026tog}, where random-matrix signatures such as the spectral-form-factor ramp can be reproduced from semiclassical gravitational contributions \cite{Saad:2018bqo}.

Recently, free probability has emerged as a natural language for describing
several aspects of quantum chaotic dynamics \cite{Jindal:2024zcg}. Free probability is the
theory of probability for non-commuting random variables \cite{SpeicherNica2006}, and is
therefore naturally adapted to quantum systems, where the algebra of observables
is non-commutative. In the context of thermalization, it plays an important role
in the generalized, or full, Eigenstate Thermalization Hypothesis, which
extends the standard ETH framework to higher-order correlations \cite{Foini:2018sdb, Pappalardi:2022aaz}. More
broadly, free probability also suggests a possible foundation for a genuinely
quantum ergodic theory. In classical ergodic theory, one classifies
dynamical systems according to the degree to which future events become
statistically independent from past events, with chaotic systems exhibiting
stronger forms of statistical independence \cite{Gesteau:2023rrx, Ouseph:2023juq}. In quantum mechanics, the
appropriate analogue of statistical independence for non-commuting observables
is freeness. It is therefore natural to classify quantum dynamics according to
the extent to which time evolution generates freeness between observables \cite{Camargo:2025zxr}.

So far, the free-probabilistic approach to quantum chaos has mostly been
developed in the context of lattice models. Recent works have used freeness to
characterize ETH and chaotic dynamics in quantum many-body systems
\cite{Fava:2023pac, Pappalardi:2023nsj,Vallini:2024bwp,Fritzsch:2024qjn, Basu:2025ubf, Pathak:2025sys, Vardhan:2025rky, Alves:2025jzl, Gill:2025upp, Fritzsch:2025arx, Dowling:2025cxr, Vallini:2025vvq, Altland:2025ngq, Herrmann:2026asu, Dowling:2026jbq}. However, this perspective is also potentially important in the context
of the AdS/CFT correspondence. One example is the use of free-probabilistic
methods in discussions of the Page curve \cite{Wang:2022ots}. Another example, more directly related to quantum chaotic dynamics, is provided
by recently proposed algebraic models of gravitational scrambling
\cite{Penington:2025hrc}. In these models, the algebra generated by early-time
and late-time observables interpolates between an early-time tensor-product
structure and a late-time free-product structure, with the latter structure
first observed in Ref.~\cite{Chandrasekaran:2022eqq}. This late-time free-product structure has a
direct probabilistic interpretation: at sufficiently large time separations, the
early- and late-time algebras become freely independent.

In this work, we use free-probability techniques to investigate full ETH in the context of quantum field theories. Building on
Ref.~\cite{Murthy:2019fgs}, where the full-ETH ansatz was related to
frequency-space constraints associated with chaotic dynamics, we first
derive directional bounds on the large-frequency behavior of the full-ETH
smooth functions in continuum QFTs using the complex-time analyticity of
regulated thermal free cumulants. The complete analyticity domain reveals
a multifrequency and generally asymmetric structure that extends the
previous lattice results. We then test these bounds in thermal CFTs. In
CFT$_2$, the universal two- and three-point functions saturate the
corresponding bounds, while at fourth order we use the holographic eikonal
description of gravitational scrambling to study genuinely dynamical
correlations. We show that the gravitational Regge contribution saturates
the strongest bounds in AdS$_2$ and in spatially homogeneous
higher-dimensional sectors. For spatially resolved higher-dimensional
correlators, the transverse shockwave dynamics introduces an additional
frequency-dependent factor that cannot violate the bounds. Finally, we discuss how these
results are consistent with the emergence of freeness in the gravitational
scrambling algebra of Ref.~\cite{Penington:2025hrc}, and clarify the
connection between the full-ETH bounds and the conventional chaos bound,
which ultimately arise from the same complex-time analytic structure.

This work is organized as follows. In Sec.~\ref{sec-scramblingAlgebra}, we review the algebraic framework for gravitational scrambling and its relation to the holographic eikonal description of OTOCs used later in the paper. In Sec.~\ref{sec:freeprobability}, we introduce the basic ingredients of free probability and free cumulants. In Sec.~\ref{sec-fullETH}, we review the full ETH ansatz and derive constraints on the large-frequency behavior of the corresponding ETH smooth functions, first for lattice systems and then for continuum QFTs. In Sec.~\ref{sec:CFTs}, we test these bounds in thermal CFTs. We first show that the universal two- and three-point functions in CFT$_2$ saturate the corresponding bounds, and then analyze the fourth-order free cumulant using the holographic eikonal description of gravitational scrambling, including its lower-dimensional and higher-dimensional extensions. We conclude in Sec.~\ref{sec-conclusions} with a discussion of the implications of our results and future directions.

\paragraph{Note added.}
While this work was being completed, Ref.~\cite{Chen:2026boh} appeared and, among other topics, studied the multitime Fourier transform of an OTOC in the de Sitter static patch. Apart from this limited overlap, the two works address substantially different questions.

\section{Algebraic Framework for Gravitational Scrambling} \label{sec-scramblingAlgebra}

In this section, we develop the algebraic framework needed to connect gravitational scrambling with the frequency-space analysis later in the paper. We begin with modular theory and half-sided modular inclusions, explaining how they encode translations along black hole horizons. We then introduce the modular-twisted product and examine how its correlators interpolate between tensor-product factorization and free independence. Finally, we connect the twisted four-point function to the holographic eikonal prescription and obtain integral representations for the lower- and higher-dimensional correlators used in our analysis of full ETH.

\subsection{Tomita--Takesaki: a brief reminder} 
A von Neumann algebra $\mathcal{A}$ acting on a Hilbert space $\mathcal{H}$ is a unital *-subalgebra of bounded operators $\mathcal{B}(\mathcal{H})$ that is closed in the weak operator topology\footnote{There are several equivalent characterizations of a von Neumann algebra. For a subalgebra of $\mathcal{B}(\mathcal{H})$ containing the identity and closed under Hermitian conjugation, closure in the weak operator topology is equivalent to closure in the strong operator topology; either condition is also equivalent to equality with its bicommutant. These equivalences concern the algebra’s closure, not equality of the topologies themselves. Closure in operator norm alone defines a $C^*$-algebra and is insufficient to ensure that it is a von Neumann algebra.}. Equivalently, by von Neumann’s double commutant theorem, $\mathcal{A}$ coincides with its double commutant $\mathcal{A}'' = (\mathcal{A}')'$, where the commutant is defined as the set of all bounded operators that commute with every element of $\mathcal{A}$. The commutant $\mathcal{A}'$ is itself a von Neumann algebra. The center of $\mathcal{A}$ is the abelian algebra denoted $\mathcal{Z}(\mathcal{A}) = \mathcal{A} \cap \mathcal{A}'$. When the center is trivial (i.e. $\mathcal{Z}(\mathcal{A}) = \mathbb{C}$), $\mathcal{A}$ is called a factor. Von Neumann algebras are classified into different types I, II$_1$, II$_\infty$, and III according to the structure of their projections and the existence or absence of normal semi-finite traces. In algebraic quantum field theory the local observable algebras are typically type III$_1$ factors; this lack of a trace is responsible for many distinctive features of entanglement and modular structure. 

Let $\omega$ be a faithful normal state on a von Neumann algebra $\mathcal{A}$. The algebraic approach trades global information of a quantum system for the local algebra of observables in a particular state. The GNS construction gives a canonical way to build the Hilbert space from the state itself. One begins with the algebra $\mathcal{A}$ viewed as a vector space and defines an inner product by $\braket{x,y} = \omega(y^{\dagger} x)$. Any element with vanishing norm corresponds to an operator that gives zero expectation value in the state for every observable; these are quotiented out. Completing the resulting pre-Hilbert space yields a Hilbert space $\mathcal{H}$. The algebra acts on this space by left multiplication, giving a representation $\pi$. The image of the unit element in $\mathcal{A}$ becomes a vector $\ket{\Omega}$ in $\mathcal{H}$ that is cyclic (the action of the algebra on it is dense in $\mathcal{H}$) and separating (the only operator that annihilates it is the zero operator). Because $\omega$ is faithful the representation is also faithful, and the original state is recovered simply as the expectation value in this vector $\omega(x) = \bra{\Omega} x \ket{\Omega}$. This is the natural Hilbert space in which to discuss entanglement, modular flow, and other state-dependent properties.

On this Hilbert space one can introduce the Tomita operator, which captures how the algebra acts differently on “kets” and “bras” with respect to the state $\ket{\Omega}$. It is first defined on the dense domain of vectors of the form $x \ket{\Omega}$ by the rule $S_0 (x \ket{\Omega}) = x^{\dagger} \ket{\Omega}$. This map is anti-linear and closable. Its closure $S$ admits a polar decomposition $S = J \Delta^{1/2}$, where $J$ is an anti-unitary operator (the modular conjugation) and $\Delta$ is a positive self-adjoint operator (the modular operator). Both operators leave the cyclic vector invariant, $J \ket{\Omega} = \ket{\Omega}$ and $\Delta \ket{\Omega} = \ket{\Omega}$. The operator $J$ implements a duality that interchanges the algebra $\mathcal{A}$ with its commutant $\mathcal{A}'$, while $\Delta$ generates a one-parameter group of automorphisms of $\mathcal{A}$ that can be interpreted as a state-dependent notion of time evolution. Together with the polar decomposition, this leads to the fundamental result of the theory:

\begin{theorem}{(Tomita-Takesaki \cite{Takesaki:1970}).}
Let $\cA$ be a von Neumann algebra with a cyclic and separating vector $\ket{\psi}$. Then $J \ket{\psi}=\ket{\psi}=\Delta \ket{\psi}$ and the following equalities hold:
\begin{equation*}
J \cA J = \cA'~, \quad {\rm and} \quad \Delta^{it} \cA \Delta^{-it} = \cA~, \quad \forall ~ t \in \mathbb{R}~.
\end{equation*}
\end{theorem}

This theorem equips the algebra-state pair with a canonical modular flow. We now analyze how this flow acts on subalgebras and how their inclusion structure encodes translations. 

\subsection{Half-sided modular inclusions}
We consider a von Neumann algebra $\mathcal{A}$ acting on a Hilbert space $\mathcal{H}$, with a cyclic and separating vector $\Omega$. We take a von Neumann subalgebra $\widetilde{\mathcal{A}} \subseteq \mathcal{A}$ for which $\Omega$ is also cyclic and separating. We denote the corresponding modular operators by $\Delta_{\widetilde{\mathcal{A}}}$ and $\Delta_{\mathcal{A}}$ and the modular conjugation of $\mathcal{A}$ by $J_{\mathcal{A}}$. While the modular flow of $\mathcal{A}$ preserves the full algebra, it may map $\widetilde{\mathcal{A}}$ into itself only for one sign of the modular parameter. We call the inclusion positive half-sided modular if
\be
\Delta_{\mathcal{A}}^{-it}\widetilde{\mathcal{A}}\Delta_{\mathcal{A}}^{it} \subseteq \widetilde{\mathcal{A}}~, \qquad \text{for all } t\ge 0 \quad.
\ee
Reversing the sign of the modular parameter defines a negative half-sided modular inclusion.

To understand how this structure encodes translations, we first recall Borchers’ theorem. We take a strongly continuous unitary group $U(s)=e^{is P}$ with a positive generator $P$, and assume that it leaves $\Omega$ invariant and maps $\mathcal{A}$ into itself for $s \geq0$, that is
\be
P\geq 0 ~,\qquad U(s)\Omega=\Omega~,
\qquad
U(s)\mathcal A U^{\dagger}(s)\subseteq\mathcal A
\quad(s\geq0)~.
\ee
Borchers' theorem then fixes its relation to the modular data
\be
\Delta_{\mathcal A}^{it}U(s)\Delta_{\mathcal A}^{-it}
=U(e^{-2\pi t}s)~,
\qquad
J_{\mathcal A}U(s)J_{\mathcal A}=U(-s)~.
\ee
Modular flow therefore rescales the translation parameter, just as a dilation rescales a displacement.

The half-sided modular inclusion theorem lets us proceed in the opposite direction. Starting from the inclusion, we can construct a positive
\emph{modular translation generator} from the two modular operators
\be
2\pi P = \log\Delta_{\tilde{\mathcal{A}}} - \log\Delta_{\mathcal{A}}~.
\ee
Here $P$ denotes the self-adjoint closure of the operator difference, and we understand the commutators below on a common domain where they are well defined.

The theorem guarantees that $P \geq0$ and that $U(s) =e^{isP}$ leaves the state invariant and satisfies
\be
\widetilde{\mathcal A}=U(1)\mathcal A U^{\dagger}(1)~.
\ee
Thus, we recover the smaller algebra by translating the larger one. This result is due to Wiesbrock \cite{Wiesbrock:1992mg}, with the proof given by Araki and Zsidó \cite{Araki:2005}. The generator $P$ also satisfies the following commutation relation
\be
[\log\Delta_{\mathcal{A}}, P] = 2\pi i P~.
\ee
If we introduce the modular Hamiltonian $K_{\mathcal{A}}=-\log \Delta_{\mathcal{A}}$, we can equivalently write 
\be
[K_{\mathcal A},P]=-2\pi iP,
\qquad
e^{itK_{\mathcal A}}Pe^{-itK_{\mathcal A}}
=e^{2\pi t}P~.
\ee

We can interpret these relations geometrically when modular flow implements a spacetime symmetry. For the vacuum algebra of a Rindler wedge, it generates Lorentz boosts. Taking a null translated subwedge gives a half-sided modular inclusion, and the associated operator $P$ generates translations along the corresponding horizon. 

Modular theory also enters the algebraic description of gravitational entropy. In Witten’s construction \cite{Witten:2021unn}, we take the crossed product of the type $\mathrm{III}_1$ algebra that emerges at large $N$ \cite{Leutheusser:2021frk} by its modular automorphism group. The resulting type $\mathrm{II}_\infty$ algebra admits a trace that allows us to define entropy up to a state-independent additive constant. For semiclassical holographic states, this algebraic entropy reproduces generalized entropy; related works include \cite{Chandrasekaran:2022eqq, Chandrasekaran:2022cip, Jensen:2023yxy, Faulkner:2024gst, Kudler-Flam:2024psh,Kudler-Flam:2023qfl,Gomez:2022eui,Speranza:2025joj, Engelhardt:2023xer, Klinger:2023auu, Klinger:2023tgi,AliAhmad:2023etg,Soni:2023fke,Aguilar-Gutierrez:2023odp,Chen:2025tbh,Espindola:2026ekv,Espindola:2026uqa, Chen:2026zxc}.

\subsection{Gravitational Scrambling Algebra}
We now use half-sided modular inclusions to describe the interaction between early- and late-time observables in gravitational scrambling. Following \cite{Penington:2025hrc}, we work in the semiclassical large-$N$ limit around a thermofield double state $\vert\Psi\rangle$. We begin with two commuting von Neumann algebras $\mathcal{A}$ and $\mathcal{B}$, acting on the tensor-product GNS Hilbert space $\mathcal{H}=\mathcal{H}_A\otimes\mathcal{H}_B$, with reference vector $\ket\Psi=\ket\Psi_A\otimes \ket\Psi_B$. These provide the untwisted ingredients of the construction. Physically, $\mathcal{A}$ is generated by single-trace boundary operators at early times $t$, with $t/\beta$ held fixed as $N\to\infty$, while $\mathcal{B}$ describes operators at late times $T(N)+t'$, where $T(N)/\beta\to\infty$ and $t'/\beta$ remains fixed.

We take the subalgebra $\widetilde{\mathcal{A}}\subset\mathcal{A}$ generated by operators with $t>0$, and the subalgebra $\tilde{\mathcal{B}}\subset\mathcal{B}$ generated by operators with $t'<0$. These naturally define positive and negative half-sided modular inclusions, respectively. We denote their positive modular translation generators by $P_A$ and $P_B$. Their opposite orientations imply
\be
[\log\Delta_A,P_A]=2\pi iP_A~,
\qquad
[\log\Delta_B,P_B]=-2\pi iP_B~.
\ee
Consequently, $P_AP_B$ is invariant under the combined modular flow generated by $\log\Delta_A+\log\Delta_B$. This property allows the interaction encoded by the modular twist to preserve the factorized modular flow.

In the gravitational bulk dual, $P_A$ and $P_B$ are identified with the integrated null energies of left- and right-moving shockwaves along the black hole and white hole horizons, respectively. When the time separation $T(N)$ is tuned to exactly the scrambling time $t_{\text{scr}} = \frac{\beta}{2\pi}\log S_{\text{BH}}$, the eikonal gravitational scattering between the two sets of modes produces a finite phase factor $e^{i P_A P_B}$. The \emph{modular-twisted product} algebra $\mathcal{R}$ is then defined as the von Neumann algebra generated by
\be
\mathcal{A}_R = e^{i P_A P_B /2}\, \mathcal{A}\, e^{-i P_A P_B /2}, \qquad 
\mathcal{B}_R = e^{-i P_A P_B /2}\, \mathcal{B}\, e^{i P_A P_B /2}.
\ee
These two subalgebras do not commute with one another; instead they obey the characteristic relation
\be
\bigl[ \mathcal{A}_R,\; e^{i P_A P_B} \mathcal{B}_R e^{-i P_A P_B} \bigr] = 0.
\ee
Importantly, $\mathcal{R}$ is a Type $\mathrm{III}_1$ factor whose commutant is the left boundary algebra $\mathcal{L}$, and its modular operator factorises as $\Delta_\Psi = \Delta_A\otimes\Delta_B$. Notably, this construction smoothly interpolates between the tensor product algebra obtained when $T(N)\ll t_{\text{scr}}$ and the free product algebra obtained when $T(N)\gg t_{\text{scr}}$, providing a unified algebraic description of gravitational scrambling.

\subsection{Twisted correlator}
The physical content of the modular-twisted product is encoded in its correlation functions. For operators $a_i\in\mathcal{A}$ and $b_i\in\mathcal{B}$ (in the factorized Hilbert space), the corresponding right-boundary operators are
\be
a_{R,i} = e^{i P_A P_B /2} a_i e^{-i P_A P_B /2}, \qquad 
b_{R,i} = e^{-i P_A P_B /2} b_i e^{i P_A P_B /2}.
\ee
Consider a time separation $t$ between the early and late operators. By evolving the $b$ operators forward in modular time (or equivalently the $a$ operators backward), one obtains the \emph{twisted product correlator}
\be \label{eq:twisted-correlator}
C(t) = \langle \Psi | a_{R,1} b_{R,1}(t) a_{R,2} b_{R,2}(t) \dots |\Psi\rangle 
= \langle \Psi_A|\langle \Psi_B|\, a_1 e^{-i\alpha P_A P_B} b_1 e^{i\alpha P_A P_B} a_2 \dots |\Psi_A\rangle|\Psi_B\rangle,
\ee
where $\alpha = e^{2\pi t}$.

This expression elegantly interpolates between two limits. In the tensor product limit $t\to -\infty$ (so that $\alpha\to 0$), the phases disappear and the correlator factorizes as $\langle a_1 a_2\dots\rangle \langle b_1 b_2\dots\rangle$, reducing the algebra to a direct product $\mathcal{A}\otimes\mathcal{B}$. In the free product limit $t\to +\infty$ (where $\alpha\to\infty$), the phases oscillate rapidly and all out-of-time-order contributions vanish unless the operators have non-zero one-point functions, yielding the free product algebra $\mathcal{A} * \mathcal{B}$. Expanding the gravitational twist near the tensor product limit, $\alpha \ll1$, produces a leading perturbative correction of order $\alpha=e^{2\pi t}$. This growth corresponds to the maximal Lyapunov exponent $\lambda_L =2\pi/\beta$ in physical time, connecting the modular inclusion structure to the saturation of the chaos bound.

\subsection{Relation to the holographic eikonal prescription}

We now summarize the relation between the modular-twisted correlator and the standard holographic representation of out-of-time-order correlators. Specializing Eq.~\eqref{eq:twisted-correlator} to a four-point function, we obtain
\begin{equation}
C_4(\vec t)
=
\langle\Psi_A,\Psi_B|
(V_1\otimes W_2)\,
S_\alpha\,
(V_3\otimes W_4)
|\Psi_A,\Psi_B\rangle\,,
\qquad
S_\alpha=e^{i\,\alpha\,P_A\,P_B}\,.
\label{eq:twisted-C4}
\end{equation}
Here $V_i$ and $W_i$ denote the untwisted representatives acting on the two tensor factors, while $P_A$ and $P_B$ are the positive generators of the corresponding half-sided modular translations. At this stage, Eq.~\eqref{eq:twisted-C4} is a purely algebraic statement and does not require a bulk interpretation.

We now invoke holography. In the semiclassical near-horizon regime, $P_A$ and $P_B$ are identified with the null-translation generators on the two horizons, while their spectral parameters $p$ and $q$ are identified with the corresponding Kruskal momenta. Under these identifications, $S_\alpha$ becomes the gravitational eikonal scattering matrix, with phase shift $\delta(p,q)=\alpha\,p\,q$. Resolving Eq.~\eqref{eq:twisted-C4} in the spectral representations of $P_A$ and $P_B$, and identifying the resulting spectral amplitudes with horizon wavefunctions, yields the standard holographic eikonal representation. This agreement is natural, since the modular-twisted product was constructed precisely to describe semiclassical gravitational scattering at time separations of order the scrambling time. A detailed derivation of this correspondence, including the precise operator dictionary and its higher-dimensional implementation, will be presented in Ref.~\cite{LevViktorHugoToAppear}.

Consider an OTOC in which the two $V$ operators are inserted at
$\mathbf{x}_1$, while the two $W$ operators are inserted at
$\mathbf{x}_2$.
\begin{equation}
    F_4(\vec z;\mathbf{x}_1,\mathbf{x}_2)
    =
    \left\langle
        V(z_1,\mathbf{x}_1)
        W(z_2,\mathbf{x}_2)
        V(z_3,\mathbf{x}_1)
        W(z_4,\mathbf{x}_2)
    \right\rangle,
    \qquad
    z_i=\frac{2\pi}{\beta}t_i+i\epsilon_i .
\end{equation}
To compute this thermal correlator holographically, it is convenient to work in a $(d+1)$-dimensional two-sided eternal asymptotically AdS black-hole geometry. In Kruskal--Szekeres coordinates, the metric can be written as
\begin{equation}
ds^2
=
-a(u,v)dudv
+
r^2(u,v)\,
d\Sigma_{d-1}^{2}(\mathbf{x})\,,
\end{equation}
where $\mathbf{x}$ denotes the coordinates on the transverse spatial geometry $\Sigma_{d-1}$, while $u$ and $v$ are Kruskal--Szekeres coordinates. The two components of the horizon are located at $u=0$ and $v=0$. With the normalization of the Kruskal coordinates adopted below, the two asymptotically AdS boundaries are located at $uv=-1$, while the future and past interior endpoints are located at $uv=1$. For later convenience, we define
$
a_0\equiv a(0)$ and 
$r_0\equiv r(0)$
which denote the horizon values of the corresponding metric functions.

The standard holographic prescription states that this correlator can be computed on the gravity side as follows
\begin{equation}
F_4(\vec z;\mathbf{x}_1,\mathbf{x}_2)
=
C_0
\int d\mathbf{x}\,d\mathbf{y}
\int_0^\infty dp\,dq\,p\,\psi_1^*(p,\mathbf{x})\psi_3(p,\mathbf{x})
\,q\,\psi_2^*(q,\mathbf{y})\,\psi_4(q,\mathbf{y})\,e^{i\,\delta(s,b)}\,,
\label{eq:eikonal-wavefunction-overlap}
\end{equation}
where $C_0 =a_0^4\,r_0^{2(d-1)}/(4\pi)^2$ and the four horizon wavefunctions are
\begin{align}
\label{eq:horizon-wavefunctions}
\psi_1(p,\mathbf{x})
&=
\int_{-\infty}^{\infty}dv\,
e^{i a_0 p v/2}\,
\left\langle
\Phi_V(0,v,\mathbf{x})\,
V^\dagger(z_1,\mathbf{x}_1)
\right\rangle_{\beta}\,,
 \nonumber
\\
\psi_3(p,\mathbf{x})
&=
\int_{-\infty}^{\infty}dv\,
e^{ia_0 p v/2}\,
\left\langle
\Phi_V(0,v,\mathbf{x})\,
V(z_3,\mathbf{x}_1)
\right\rangle_{\beta}\,,
 \nonumber
\\
\psi_2(q,\mathbf{y})
&=
\int_{-\infty}^{\infty}du\,
e^{i a_0 q u/2}\,
\left\langle
\Phi_W(u,0,\mathbf{y})\,
W^\dagger(z_2,\mathbf{x}_2)
\right\rangle_{\beta}\,,
 \nonumber
\\
\psi_4(q,\mathbf{y})
&=
\int_{-\infty}^{\infty}du\,
e^{i a_0 q u/2}\,
\left\langle
\Phi_W(u,0,\mathbf{y})\,
W(z_4,\mathbf{x}_2)
\right\rangle_{\beta}\,.
\end{align}
Here $p\equiv p^u>0$ and $q\equiv p^v>0$ are the null Kruskal momenta associated with the $u=0$ and $v=0$ horizons, respectively. The eikonal phase is given by
\begin{equation}
\delta(p,q) = g \,a_0\,p\,q\, h(\mathbf{x},\mathbf{y})~,
\end{equation}
where $g= 4 \pi \,G_N/r_0^{d-3}$ and the transverse shock-wave profile is determined by Einstein's equations and is the Green's function of a massive Laplacian on the transverse horizon geometry,
\begin{equation}
\left(
-\nabla_{\perp}^{2}
+
\mu^2
\right)
h(\mathbf{x},\mathbf{y})
=
\frac{
\delta^{(d-1)}(\mathbf{x}-\mathbf{y})
}{
\sqrt{g_{\perp}(\mathbf{x})}
}\,.
\label{eq}
\end{equation}

\paragraph{Results for Rindler-AdS$_{d+1}$/CFT$_d$.}
We now specialize to Rindler-AdS$_{d+1}$, whose transverse geometry is the unit hyperbolic space $H^{d-1}$, and $a_0=4$. The corresponding OTOCs were derived in Ref.~\cite{Ahn:2019rnq}. Here, we keep all four insertion times arbitrary. We denote by $d(\mathbf{x},\mathbf{y})$ the geodesic distance between two points in $H^{d-1}$, measured in units of the AdS radius. By hyperbolic symmetry, the transverse shock-wave profile depends only on this distance and, at large separation, behaves as
$h(\mathbf{x},\mathbf{y})
\sim
e^{
-(d-1)
d(\mathbf{x},\mathbf{y})}
$ for $d(\mathbf{x},\mathbf{y})\gg 1\,.$ On the horizons, the bulk-to-boundary propagator takes the form
\begin{equation}
    K_\Delta(u,v,\mathbf{x}|z_i,\mathbf{x}_i)
    =
    c_\Delta
    \left[
        u\,e^{z_i}
        -
        v\,e^{-z_i}
        +
        (1-uv)
        \cosh d(\mathbf{x},\mathbf{x}_i)
    \right]^{-\Delta},
\end{equation} 
Fourier transforming this propagator along the two horizons and rescaling the null momenta
gives the integral expression
\begin{align}
    F_4(\vec z;\mathbf{x}_1,\mathbf{x}_2)
    &=
    {\cal N}\,
    {\cal P}_{13}^{(V)}(\vec z)\,
    {\cal P}_{24}^{(W)}(\vec z)
    \int_{H^{d-1}}d\mathbf{x}\,d\mathbf{y}
    \int_0^\infty dp\,dq\,
    p^{2\Delta_V-1}q^{2\Delta_W-1}
    \nonumber\\
    &\quad\times
    e^{-p\cosh d(\mathbf{x},\mathbf{x}_1)}
    e^{-q\cosh d(\mathbf{y},\mathbf{x}_2)}
    \exp\left[
        i\, g \,a_0 \,pq\,
        h\bigl(\mathbf{x},\mathbf{y}\bigr)
        {\cal R}(\vec z)
    \right],
    \label{eq:higher-dimensional-resolved-otoc}
\end{align}
where $\mathcal N =
\left[
\frac{
4\pi\,c_{\Delta_V}\,c_{\Delta_W}
}{
\Gamma(\Delta_V)\,
\Gamma(\Delta_W)
}
\right]^2
$ and
\begin{equation}
    {\cal P}_{13}^{(V)}(\vec z)
    =
    \left[
        2\sinh\left(\frac{z_3-z_1}{2}\right)
    \right]^{-2\Delta_V},
    \qquad
    {\cal P}_{24}^{(W)}(\vec z)
    =
    \left[
        2\sinh\left(\frac{z_4-z_2}{2}\right)
    \right]^{-2\Delta_W},
\end{equation}
while
\begin{equation}
    {\cal R}(\vec z)
    =
    \frac{1}{
        \left(e^{z_3}-e^{z_1}\right)
        \left(e^{-z_4}-e^{-z_2}\right)
    }.
    \label{eq:resolved-regge-factor-higher-d}
\end{equation}

\paragraph{Lower-dimensional case.} Notice that the corresponding result for AdS$_2$/CFT$_1$ derived in \cite{Maldacena:2016upp} can be obtained from the above result by setting the transverse profile to unity, omitting the transverse spatial integrals, and replacing all factors of $\cosh d$ by one. This gives
\begin{equation}
F_4(\vec z)
=
\frac{
\mathcal P_{13}^{(V)}(\vec z)\,
\mathcal P_{24}^{(W)}(\vec z)
}{
\Gamma(2\Delta_V)\,
\Gamma(2\Delta_W)
}
\int_0^\infty
dp\,dq\,
p^{2\Delta_V-1}\,
q^{2\Delta_W-1}\,
e^{
-p-q
+i
g a_0 p q\,\mathcal R(\vec z)}\,.
\label{eq:AdS2-eikonal-result}
\end{equation}

\section{Free Probability and Free Cumulants}
\label{sec:freeprobability}
In this section, we provide a brief review of free probability theory, introducing the main concepts and tools relevant for the discussion of the full ETH. More comprehensive and pedagogical introductions to the subject can be found in \cite{SpeicherNica2006,novak2012lecturesfreeprobability,speicher2019lecture}. Additional pedagogical lecture notes emphasizing applications to quantum chaos and ETH can be found in \cite{PappalardiLectureNotes}. For a concise review aimed at the high-energy theory community, see \cite{Wang:2022ots}. 

Free probability originated in the study of von Neumann algebras. Voiculescu introduced the theory to investigate the structure of type $\mathrm{II}_1$ factors, particularly those associated with free groups \cite{10.1007/BFb0074909, VOICULESCU2005127}. In this framework, the normalized trace serves as an expectation value, while free independence captures the probabilistic structure of free products. The theory subsequently developed important applications to subfactor theory \cite{guionnet2008}.

More generally, free probability theory extends classical probability theory (see App.~\ref{app:ClassicalProbability} for a brief review) to situations in which the random variables do not commute. The first ingredient is the notion of a non-commutative probability space.

\paragraph{Definition I. Non-commutative probability space.}
A non-commutative probability space is a pair $(\mathcal{A},\varphi)$, where $\mathcal{A}$ is a unital von Neumann algebra and $\varphi:\mathcal{A}\rightarrow\mathbb{C}$ is a unital, positive, and faithful linear functional.

Intuitively, $\mathcal{A}$ plays the role of the algebra of random variables, while $\varphi$ generalizes the notion of expectation value. An important example is provided by quantum mechanics, where $\mathcal{A}\subset B(\mathcal{H})$ is a subalgebra of bounded operators acting on a Hilbert space $\mathcal{H}$, and $\varphi$ is a state assigning expectation values to observables. In situations where a trace representation exists, one may write
\begin{equation}
\varphi(a)=\mathrm{Tr}(\rho \,a),
\end{equation}
where $\rho$ is a density matrix. Particular cases include pure states, for which $\rho=|\psi\rangle\langle\psi|$\footnote{Here we illustrate states in general, without requiring faithfulness. A pure state on $\mathcal B(\mathcal H)$, with $\dim\mathcal H>1$, assigns zero expectation value to nonzero projections orthogonal to $\ket \psi$, so it does not satisfy the additional faithfulness assumption in Definition I.} and $\varphi(a)=\langle\psi|a|\psi\rangle$, and thermal (Gibbs) states, for which $\rho=e^{-\beta H}/Z$ with $Z=\mathrm{Tr}(e^{-\beta H})$. From now on, we will use the shorthand notation
\begin{equation}
\langle a\rangle \equiv \varphi(a)
\end{equation}
to denote expectation values.

In classical probability spaces, two random variables $X$ and $Y$ are independent if the events they generate are independent, which implies factorization of mixed moments, $\mathbb{E}[X^nY^m]=\mathbb{E}[X^n]\mathbb{E}[Y^m]$, or, equivalently, the vanishing of mixed cumulants involving such variables, $c_n(X,\dots,X,Y,\dots,Y)=0$. It is possible to define a similar notion of independence in non-commutative probability spaces, with the caveat that random variables do not commute and therefore their ordering inside expectation values matters. Such notion is called \textit{freeness}, or \textit{free independence}, and is defined as follows.

\paragraph{Definition II. Free independence.}
Two random variables $a,b\in\mathcal A$ are said to be freely independent, or simply free, if
\begin{equation}
\left\langle
\left(p_1(a)-\langle p_1(a)\rangle\right)
\left(p_2(b)-\langle p_2(b)\rangle\right)
\cdots
\left(p_n(b)-\langle p_n(b)\rangle\right)
\right\rangle =0,
\end{equation}
for arbitrary polynomials $p_j$, where successive factors alternate between polynomials in $a$ and polynomials in $b$, and $n\geq1$.

For simplicity, let us consider the special case in which $p_i(x)=x$ for all $i=1,\ldots,n$. In this case, the definition reduces to
\begin{equation}
\left\langle
\prod_{j=1}^{n}
\left(
x_j-\langle x_j\rangle
\right)
\right\rangle=0,
\end{equation}
where $x_j\in\{a,b\}$ and adjacent factors satisfy $x_j\neq x_{j+1}$.

Applying this condition recursively for increasing values of $n$, one obtains, for example,
\begin{align}
\langle ab\rangle 
&=
\langle a\rangle\langle b\rangle,\nonumber
\\
\langle aba\rangle
&=
\langle a^2\rangle\langle b\rangle,\nonumber
\\
\langle abab\rangle
&=
\langle a^2\rangle\langle b\rangle^2
+
\langle b^2\rangle\langle a\rangle^2
-
\langle a\rangle^2\langle b\rangle^2 .
\end{align}

Additionally, taking $p_i(x)=x^2$ for all $i=1,\dots,n$ and $n=2$, one obtains
\begin{equation}
\langle aabb\rangle=\langle a^2\rangle \langle b^2\rangle .
\end{equation}
Therefore, freeness should be understood as a notion of independence in the sense that joint expectation values, or mixed moments, involving both $a$ and $b$ are entirely determined by moments involving only $a$ or only $b$. The previous definition can be generalized from random variables to algebras as follows.

\paragraph{Definition III. Free independence of algebras.}
A collection of unital subalgebras $\{\mathcal A_i\}_{i\in I}\subset\mathcal A$ is said to be freely independent, or free, if
\begin{equation}
\left\langle a_1a_2\cdots a_n\right\rangle=0,
\end{equation}
whenever $a_j\in\mathcal A_{i_j}$ with adjacent indices satisfying $i_j\neq i_{j+1}$ and $\langle a_j\rangle=0$ for all $j=1,\ldots,n$.

One can also define freeness equivalently using the concept of free cumulants, which are defined as follows. 

\paragraph{Definition IV. Free cumulants.}
The free cumulants associated with random variables $a_1,\dots,a_n$ are defined recursively through the moment expansion
\begin{equation} \label{eq:momentfromfreecumulants1}
\left\langle a_1a_2\cdots a_n\right\rangle
=
\sum_{\pi\in NC(n)}
\prod_{B\in\pi}
\kappa_{|B|}
\left(
a_{B(1)},a_{B(2)},\dots,a_{B(|B|)}
\right),
\end{equation}
where $NC(n)$ denotes the set of non-crossing partitions of $\{1,\dots,n\}$, the product runs over all blocks $B$ of the partition $\pi$, and $\kappa_{|B|}$ denotes the free cumulant associated with the variables whose indices belong to the block $B$.

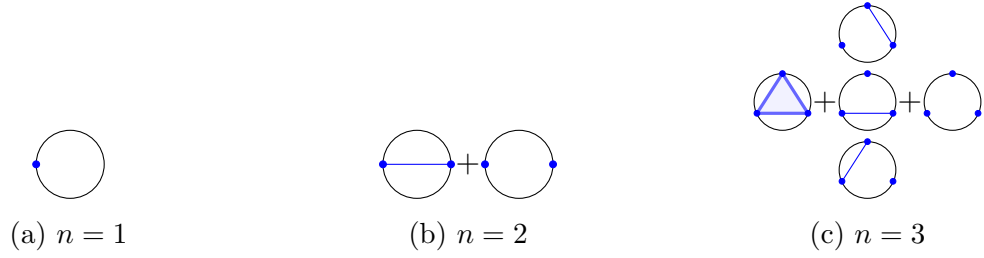
\begin{figure}[h!]
    \centering

    \begin{subfigure}[b]{0.32\textwidth}
        \centering
        \begin{tikzpicture}[scale=0.9]
            \draw (0,0) circle(0.5);
            \draw[fill=blue,blue](-0.5,0) circle [radius=0.05];
        \end{tikzpicture}
        \caption{$n=1$ }
        
    \end{subfigure}
    \hfill
    \begin{subfigure}[b]{0.32\textwidth}
        \centering
        \begin{tikzpicture}[scale=0.9]
            \draw (-0.5,0) circle(0.5);
            \draw[fill=blue,blue](-1,0) circle [radius=0.05];
            \draw[fill=blue,blue](0,0) circle [radius=0.05];
            \draw[blue] (-1,0)--(0,0);
            \node at (0.25,0){$+$};
            \draw (1,0) circle(0.5);
            \draw[fill=blue,blue](0.5,0) circle [radius=0.05];
            \draw[fill=blue,blue](1.5,0) circle [radius=0.05];
        \end{tikzpicture}
        \caption{$n=2$}
       
    \end{subfigure}
    \hfill
    \begin{subfigure}[b]{0.32\textwidth}
        \centering
        \begin{tikzpicture}[scale=0.75]
            \draw (-0.5,0) circle(0.5);
            \filldraw[color=blue!60, fill=blue!5, very thick]
                (-0.95,-0.2) -- (-0.05,-0.2) -- (-0.5,0.5) -- cycle;
            \draw[fill=blue,blue](-0.95,-0.2) circle [radius=0.05];
            \draw[fill=blue,blue](-0.05,-0.2) circle [radius=0.05];
            \draw[fill=blue,blue](-0.5,0.5) circle [radius=0.05];

            \node at (0.25,0){$+$};
            
            \draw (1,0) circle(0.5);
            \draw[fill=blue,blue](1.45,-0.2) circle [radius=0.05];
            \draw[fill=blue,blue](0.55,-0.2) circle [radius=0.05];
            \draw[fill=blue,blue](1,0.5) circle [radius=0.05];
            \draw[blue] (1.45,-0.2)--(0.55,-0.2);

            \draw (1,1.2) circle(0.5);
            \draw[fill=blue,blue](1.45,1) circle [radius=0.05];
            \draw[fill=blue,blue](0.55,1) circle [radius=0.05];
            \draw[fill=blue,blue](1,1.7) circle [radius=0.05];
            \draw[blue] (1.45,1)--(1,1.7);

            \draw (1,-1.2) circle(0.5);
            \draw[fill=blue,blue](1.45,-1.4) circle [radius=0.05];
            \draw[fill=blue,blue](0.55,-1.4) circle [radius=0.05];
            \draw[fill=blue,blue](1,-0.7) circle [radius=0.05];
            \draw[blue] (0.55,-1.4)--(1,-0.7);
             
            \node at (1.75,0){$+$};

            \draw (2.5,0) circle(0.5);
            \draw[fill=blue,blue](2.95,-0.2) circle [radius=0.05];
            \draw[fill=blue,blue](2.05,-0.2) circle [radius=0.05];
            \draw[fill=blue,blue](2.5,0.5) circle [radius=0.05];
        \end{tikzpicture}
        \caption{$n=3$}
        
    \end{subfigure}

    \captionsetup{justification=raggedright,singlelinecheck=false}
    \caption{Diagrammatic representation of set partitions for small values of $n$. Panels (a), (b), and (c) show the partitions of the sets $\{1\}$, $\{1,2\}$, and $\{1,2,3\}$, respectively. In each diagram, elements are placed on a circle in cyclic order, and blocks are represented by chords connecting the corresponding points.}
    \label{fig:partitionn3}
\end{figure}

Let us consider a few illustrative examples:
\begin{align}
\langle a_1 \rangle &= \kappa_1(a_1)\\
\langle a_1 a_2\rangle
&=
\kappa_1(a_1)\kappa_1(a_2)
+\kappa_2(a_1,a_2),
\\
\langle a_1a_2a_3\rangle
&=
\kappa_1(a_1)\kappa_1(a_2)\kappa_1(a_3)
+\kappa_2(a_1,a_2)\kappa_1(a_3)
+\kappa_1(a_1)\kappa_2(a_2,a_3)
\nonumber
\\
&\quad
+\kappa_2(a_1,a_3)\kappa_1(a_2)
+\kappa_3(a_1,a_2,a_3).
\end{align}
Up to $n=3$, the relation between moments and free cumulants in Eq.~(\ref{eq:momentfromfreecumulants1}) coincides with the corresponding relation for classical cumulants. This equivalence arises because, for $n\leq 3$, all set partitions are non-crossing, as illustrated in Fig.~\ref{fig:partitionn3}. 

\begin{figure}[h!]
    \centering
    
    \begin{subfigure}[b]{\textwidth}
    \centering
        \begin{tikzpicture}
        \centering
        \draw (-0.5,0) circle(0.5);
        \filldraw[color=blue!60, fill=blue!5, very thick](-0.9,-0.3) rectangle (-0.1,0.3);
        \draw[fill=blue,blue](-0.9,-0.3) circle [radius=0.05];
        \draw[fill=blue,blue](-0.1,-0.3) circle [radius=0.05];
        \draw[fill=blue,blue](-0.9,0.3) circle [radius=0.05];
        \draw[fill=blue,blue](-0.1,0.3) circle [radius=0.05];

        \node at (0.25,0){$+$};
        
        \draw (1,0) circle(0.5);
        \draw[fill=blue,blue](0.6,-0.3) circle [radius=0.05];
        \draw[fill=blue,blue](1.4,-0.3) circle [radius=0.05];
        \draw[fill=blue,blue](0.6,0.3) circle [radius=0.05];
        \draw[fill=blue,blue](1.4,0.3) circle [radius=0.05];
        \draw[blue] (0.6,-0.3)--(1.4,-0.3);
        \draw[blue] (0.6,0.3)--(1.4,0.3);

        \draw (1,1.2) circle(0.5);
        \draw[fill=blue,blue](0.6,0.9) circle [radius=0.05];
        \draw[fill=blue,blue](1.4,0.9) circle [radius=0.05];
        \draw[fill=blue,blue](0.6,1.5) circle [radius=0.05];
        \draw[fill=blue,blue](1.4,1.5) circle [radius=0.05];
        \draw[blue] (0.6,1.5)--(0.6,0.9);
        \draw[blue] (1.4,0.9)--(1.4,1.5);

        \draw (1,-1.2) circle(0.5);
        \draw[fill=blue,blue](0.6,-1.5) circle [radius=0.05];
        \draw[fill=blue,blue](1.4,-1.5) circle [radius=0.05];
        \draw[fill=blue,blue](0.6,-0.9) circle [radius=0.05];
        \draw[fill=blue,blue](1.4,-0.9) circle [radius=0.05];
         \draw[red] (0.6,-1.5)--(1.4,-0.9);
         \draw[red] (1.4,-1.5)--(0.6,-0.9);
         
         \node at (1.75,0){$+$};

         \draw (2.5,0) circle(0.5);
        \draw[fill=blue,blue](2.1,-0.3) circle [radius=0.05];
        \draw[fill=blue,blue](2.9,-0.3) circle [radius=0.05];
        \draw[fill=blue,blue](2.1,0.3) circle [radius=0.05];
        \draw[fill=blue,blue](2.9,0.3) circle [radius=0.05];
        \draw[blue] (2.1,0.3)--(2.1,-0.3);

        \draw (2.5,1.2) circle(0.5);
        \draw[fill=blue,blue](2.1,0.9) circle [radius=0.05];
        \draw[fill=blue,blue](2.9,0.9) circle [radius=0.05];
        \draw[fill=blue,blue](2.1,1.5) circle [radius=0.05];
        \draw[fill=blue,blue](2.9,1.5) circle [radius=0.05];
        \draw[blue] (2.9,1.5)--(2.1,1.5);

        \draw (2.5,-1.2) circle(0.5);
        \draw[fill=blue,blue](2.1,-1.5) circle [radius=0.05];
        \draw[fill=blue,blue](2.9,-1.5) circle [radius=0.05];
        \draw[fill=blue,blue](2.1,-0.9) circle [radius=0.05];
        \draw[fill=blue,blue](2.9,-0.9) circle [radius=0.05];
        \draw[blue] (2.9,-1.5)--(2.1,-1.5);

        \draw (4,0) circle(0.5);
        \draw[fill=blue,blue](3.6,-0.3) circle [radius=0.05];
        \draw[fill=blue,blue](4.4,-0.3) circle [radius=0.05];
        \draw[fill=blue,blue](3.6,0.3) circle [radius=0.05];
        \draw[fill=blue,blue](4.4,0.3) circle [radius=0.05];
        \draw[blue] (4.4,-0.3)--(4.4,0.3);

        \draw (4,1.2) circle(0.5);
        \draw[fill=blue,blue](3.6,0.9) circle [radius=0.05];
        \draw[fill=blue,blue](4.4,0.9) circle [radius=0.05];
        \draw[fill=blue,blue](3.6,1.5) circle [radius=0.05];
        \draw[fill=blue,blue](4.4,1.5) circle [radius=0.05];
        \draw[blue] (4.4,1.5)--(3.6,0.9);

        \draw (4,-1.2) circle(0.5);
        \draw[fill=blue,blue](3.6,-1.5) circle [radius=0.05];
        \draw[fill=blue,blue](4.4,-1.5) circle [radius=0.05];
        \draw[fill=blue,blue](3.6,-0.9) circle [radius=0.05];
        \draw[fill=blue,blue](4.4,-0.9) circle [radius=0.05];
        \draw[blue] (3.6,-0.9)--(4.4,-1.5);

         \node at (4.75,0){$+$};

        \draw (5.5,0.75) circle(0.5);
        \filldraw[color=blue!60, fill=blue!5, very thick](5.1,1.05) -- (5.1,0.45) -- (5.9,0.45) -- cycle;
        \draw[fill=blue,blue](5.1,0.45) circle [radius=0.05];
        \draw[fill=blue,blue](5.9,0.45) circle [radius=0.05];
        \draw[fill=blue,blue](5.1,1.05) circle [radius=0.05];
        \draw[fill=blue,blue](5.9,1.05) circle [radius=0.05];

        \draw (5.5,-0.75) circle(0.5);
        \filldraw[color=blue!60, fill=blue!5, very thick](5.9,-0.45) -- (5.9,-1.05) -- (5.1,-1.05) -- cycle;
        \draw[fill=blue,blue](5.1,-1.05) circle [radius=0.05];
        \draw[fill=blue,blue](5.9,-1.05) circle [radius=0.05];
        \draw[fill=blue,blue](5.1,-0.45) circle [radius=0.05];
        \draw[fill=blue,blue](5.9,-0.45) circle [radius=0.05];

        \draw (7,0.75) circle(0.5);
        \filldraw[color=blue!60, fill=blue!5, very thick](6.6,0.45) -- (7.4,1.05) -- (6.6,1.05) -- cycle;
        \draw[fill=blue,blue](6.6,0.45) circle [radius=0.05];
        \draw[fill=blue,blue](7.4,0.45) circle [radius=0.05];
        \draw[fill=blue,blue](6.6,1.05) circle [radius=0.05];
        \draw[fill=blue,blue](7.4,1.05) circle [radius=0.05];

        \draw (7,-0.75) circle(0.5);
        \filldraw[color=blue!60, fill=blue!5, very thick](6.6,-0.45) -- (7.4,-0.45) -- (7.4,-1.05) -- cycle;
        \draw[fill=blue,blue](6.6,-1.05) circle [radius=0.05];
        \draw[fill=blue,blue](7.4,-1.05) circle [radius=0.05];
        \draw[fill=blue,blue](6.6,-0.45) circle [radius=0.05];
        \draw[fill=blue,blue](7.4,-0.45) circle [radius=0.05];

         \node at (7.75,0){$+$};

         \draw (8.5,0) circle(0.5);
        \draw[fill=blue,blue](8.1,-0.3) circle [radius=0.05];
        \draw[fill=blue,blue](8.9,-0.3) circle [radius=0.05];
        \draw[fill=blue,blue](8.1,0.3) circle [radius=0.05];
        \draw[fill=blue,blue](8.9,0.3) circle [radius=0.05];
    
    \end{tikzpicture}
    \end{subfigure}
    \captionsetup{justification=raggedright,singlelinecheck=false}
    \caption{Diagrammatic representation of all set partitions of $\{1,2,3,4\}$, drawn on a circle with the elements ordered cyclically. Each diagram corresponds to a partition $\pi \in \mathcal P(4)$, with blocks represented by chords connecting the corresponding points. The unique crossing partition is highlighted by intersecting chords drawn in red, while all non-crossing partitions $\pi \in NC(4)$ are shown in blue.
}
    \label{fig:partitionn4}
\end{figure}
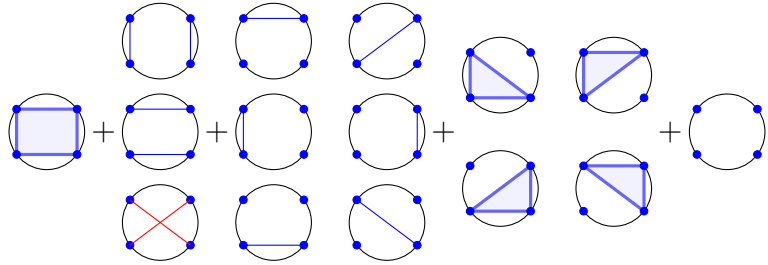

The difference between classical and free cumulants first appears at order $4$, where the first crossing partitions emerge (see Fig.~\ref{fig:partitionn4}). In this case, the fourth moment can be written in terms of free cumulants as
\begin{align*}
\langle a_1a_2a_3a_4\rangle
&=
\kappa_1(a_1)\kappa_1(a_2)\kappa_1(a_3)\kappa_1(a_4)
+\kappa_2(a_1,a_2)\kappa_1(a_3)\kappa_1(a_4)
\nonumber
\\
&\quad
+\kappa_2(a_1,a_3)\kappa_1(a_2)\kappa_1(a_4)
+\kappa_2(a_1,a_4)\kappa_1(a_2)\kappa_1(a_3)
\nonumber
\\
&\quad
+\kappa_1(a_1)\kappa_2(a_2,a_3)\kappa_1(a_4)
+\kappa_1(a_1)\kappa_2(a_2,a_4)\kappa_1(a_3)
\nonumber
\\
&\quad
+\kappa_1(a_1)\kappa_1(a_2)\kappa_2(a_3,a_4)
+\kappa_2(a_1,a_2)\kappa_2(a_3,a_4)
\nonumber
\\
&\quad
+\kappa_2(a_1,a_4)\kappa_2(a_2,a_3)
+\kappa_3(a_1,a_2,a_3)\kappa_1(a_4)
\nonumber
\\
&\quad
+\kappa_3(a_1,a_2,a_4)\kappa_1(a_3)
+\kappa_3(a_1,a_3,a_4)\kappa_1(a_2)
\nonumber
\\
&\quad
+\kappa_1(a_1)\kappa_3(a_2,a_3,a_4)
+\kappa_4(a_1,a_2,a_3,a_4).
\end{align*}
In contrast to the classical case, the crossing contribution
$\kappa_2(a_1,a_3)\kappa_2(a_2,a_4)$ is absent, since free cumulants receive contributions only from non-crossing partitions. For a clearer comparison with the classical case given in Eq.~(\ref{eq:MomentsFromCumulants}), consider the case where $a_1=a_2=a_3=a_4$, in which case one obtains
\begin{equation}
\langle a^4\rangle
=
\kappa_1^4(a)
+
6\kappa_1^2(a)\kappa_2(a)
+
2\kappa_2^2(a)
+
4\kappa_1(a)\kappa_3(a)
+
\kappa_4(a) .
\end{equation}
Here, we introduced the shorthand notation $\kappa_n(a)\equiv\kappa_n(a,\ldots,a)$ for cumulants evaluated on repeated arguments. Notice that the coefficient of $\kappa_2^2$ is $2$ for free cumulants, instead of $3$ as in the classical case. This difference reflects the absence of the crossing partition, highlighted with red chords in Fig.~\ref{fig:partitionn4}.

Now we are finally in position to state an equivalent characterization of freeness in terms of mixed free cumulants. A collection of unital subalgebras $\mathcal A_1,\mathcal A_2,\dots$ is free if all mixed free cumulants vanish, namely
\begin{equation}
\kappa_n(a_1,\dots,a_n)=0,
\end{equation}
whenever $a_j\in\mathcal A_{i_j}$ and not all indices $i_j$ are equal.

%Free probability theory is a generalization of classical probability theory (see App.~\ref{app:ClassicalProbability} for a brief review) to the case in which the random variables do not commute. As reviewed in App.~\ref{app:ClassicalProbability}, a classical probability space admits a purely algebraic formulation in terms of a pair $(\mathcal{A},\varphi)$, where $\mathcal{A}$ denotes a commutative algebra of bounded random variables and $\varphi$ is a linear functional playing the role of expectation value. The non-commutative generalization is obtained by allowing the algebra $\mathcal{A}$ to be non-commutative. In this setting, however, one must introduce a generalized notion of independence adapted to non-commuting variables, known as free independence.

%The relevance of free probability theory to physics stems from the fact that quantum mechanical observables naturally form a non-commutative algebra. In particular, free independence, which originally arose in the context of infinite-dimensional random matrices, may also emerge in many-body quantum systems in the thermodynamic limit, where the number of degrees of freedom becomes large. In this limit, correlations between suitable classes of observables may asymptotically organize according to the structures of free probability theory.

\section{Full Eigenstate Thermalization Hypothesis} \label{sec-fullETH}

In this section, we review the so-called full Eigenstate Thermalization Hypothesis (ETH) \cite{Foini:2018sdb}, which extends the standard ETH \cite{Deutsch:1991msp,Srednicki:1994mfb} to provide a consistent description of higher-order correlation functions. In particular, we present its formulation in the language of free probability theory, following \cite{Pappalardi:2022aaz}.

\subsection{Standard ETH and its limitations}

The standard Eigenstate Thermalization Hypothesis (sETH) \cite{Deutsch:1991msp,Srednicki:1994mfb} provides the leading explanation for thermalization in isolated quantum many-body systems. It postulates that the matrix elements of local operators in the energy eigenbasis take the universal form
\begin{equation}
    \mathcal{O}_{ij}\equiv\langle E_i | \mathcal{O} |E_j\rangle
    =
    \mathcal{O}(\bar{E}_{ij}) \delta_{ij}
    +e^{-S(\bar{E}_{ij})/2}
    f_{\mathcal{O}}(\bar{E}_{ij},\omega_{ij})R_{ij},
\end{equation}
where
$
\bar{E}_{ij}=\frac{E_i+E_j}{2},
\qquad
\omega_{ij}=E_i-E_j,
$
and $S(\bar{E}_{ij})$ is the thermodynamic entropy at the average energy $\bar{E}_{ij}$. The smooth function $\mathcal{O}(\bar{E}_{ij})$ coincides, up to finite-size corrections, with the microcanonical expectation value of $\mathcal{O}$ at energy $\bar{E}_{ij}$. The function $f_{\mathcal{O}}(\bar{E}_{ij},\omega_{ij})$ is also smooth and encodes the structure of equilibrium two-point correlation functions. Finally, $R_{ij}$ are assumed to be Gaussian random variables with zero mean and unit variance, describing the fluctuations of the matrix elements around their smooth envelope. Strictly speaking, unless the Hamiltonian itself is drawn from a random ensemble, the quantities $R_{ij}$ are not genuinely random variables. Rather, they provide an effective statistical description of the fluctuations of matrix elements within a sufficiently narrow microcanonical energy window, where averages are taken over nearby eigenstates.

Now we would like to explore the consequences of the standard ETH for correlation functions. Consider, for example, the thermal two-point function
\begin{equation}
    \begin{aligned}
    \langle \mathcal{O}(t)\mathcal{O}(0)\rangle_\beta
    &=
    \frac{1}{Z_\beta}\mathrm{Tr}\!\left[e^{-\beta H}\mathcal{O}(t)\mathcal{O}(0)\right] \\
    &=
    \frac{1}{Z_\beta}
    \sum_{i,j}
    e^{-\beta E_i}
    e^{i(E_i-E_j)t}
    \mathcal{O}_{ij}\mathcal{O}_{ji},
    \end{aligned}
\end{equation}
where
$
Z_\beta=\mathrm{Tr}\!\left[e^{-\beta H}\right].
$
To use the ETH ansatz as an input for this correlation function, we need to know the statistical averages of the products $\mathcal{O}_{ij}\mathcal{O}_{ji}$, which naturally separate into the cases $i=j$ and $i\neq j$.

The standard ETH can be equivalently formulated in terms of the statistical moments of the fluctuations $R_{ij}$. Denoting by an overline the average over these fluctuations within a narrow microcanonical energy window, the ETH ansatz assumes
\begin{equation} \label{eq:Rstat}
    \overline{R_{ij}}=0,
    \qquad
    \overline{R_{ij}R_{kl}}
    =
    \delta_{il}\delta_{jk},
\end{equation}
which immediately implies
\begin{equation}
    \overline{\mathcal{O}_{ij}}
    =
    \mathcal{O}(E_i)\delta_{ij},
\end{equation}
and
\begin{equation}
    \overline{\mathcal{O}_{ij}\mathcal{O}_{kl}}
    =
    \mathcal{O}(\bar E_{ij})\mathcal{O}(\bar E_{kl})
    \delta_{ij}\delta_{kl}
    +
    e^{-S(\bar E_{ij})}
    F^{(2)}_{\bar E_{ij}}(\omega_{ij})
    \delta_{il}\delta_{jk},
\end{equation}
where
\begin{equation}
    F^{(2)}_{\bar E}(\omega)
    =
    f_{\mathcal O}(\bar E,\omega)
    f_{\mathcal O}(\bar E,-\omega).
\end{equation}
Thus, the standard ETH completely determines the one- and two-point statistics of matrix elements. In particular, setting $k=j$ and $l=i$ in the previous equation yields precisely the statistical averages required in the thermal two-point function:
\begin{equation}
\overline{\mathcal{O}_{ij}\mathcal{O}_{ji}}
=
\begin{cases}
\mathcal{O}(E_i)^2, & i=j,\\[0.3em]
e^{-S(\bar E_{ij})}
F^{(2)}_{\bar E_{ij}}(\omega_{ij}), & i\neq j.
\end{cases}
\end{equation}

These expressions can now be substituted into the spectral decomposition of the thermal correlator, showing that the standard ETH completely determines equilibrium two-point functions.

Note, however, that for higher-order correlation functions of the form
\begin{equation}
\begin{aligned}
\left\langle
\mathcal{O}(t_1)\mathcal{O}(t_2)\cdots\mathcal{O}(t_n)
\right\rangle_\beta
&=
\frac{1}{Z_\beta}
\sum_{i_1,\ldots,i_n}
e^{-\beta E_{i_1}}
\,
\mathcal{O}_{i_1 i_2}
\mathcal{O}_{i_2 i_3}
\cdots
\mathcal{O}_{i_n i_1}
\\
&\hspace{1cm}\times
\exp\!\left[
i\sum_{a=1}^{n}
\omega_{i_a i_{a+1}}\,t_a
\right],
\end{aligned}
\end{equation}
where $i_{n+1}\equiv i_1$. In contrast to the two-point function, the evaluation of such correlators within the ETH framework requires the statistical averages of products of $n$ matrix elements,
\begin{equation} \label{eq:productOs}
\overline{
\mathcal{O}_{i_1 i_2}
\mathcal{O}_{i_2 i_3}
\cdots
\mathcal{O}_{i_n i_1}
}.
\end{equation}
These quantities are not determined by the standard ETH, which only specifies the first and second moments of the matrix-element distribution. This limitation motivates the full Eigenstate Thermalization Hypothesis, which provides an ansatz for the complete hierarchy of statistical moments \cite{Foini:2018sdb,Pappalardi:2022aaz}.

%One might attempt to circumvent this limitation by imposing the additional assumption that the fluctuations $R_{ij}$ are jointly Gaussian random variables satisfying Eq.~\eqref{eq:Rstat}. In that case, Wick's theorem implies that every higher-order moment factorizes into a sum over pairwise contractions, so that all higher-order correlation functions are completely determined by the two-point ETH data encoded in $F^{(2)}_{\bar E}(\omega)$. However, there is no physical justification for such a Gaussian assumption in generic quantum many-body systems, and the full ETH instead allows for genuinely independent higher-order correlations.

\subsection{Full ETH: an ansatz for higher-order correlations}

To characterize higher-order correlation functions, the full ETH postulates an ansatz for the statistical averages of products of matrix elements. For products whose indices are all distinct, one assumes
\begin{align}
\overline{
\mathcal{O}_{i_1 i_2}
\mathcal{O}_{i_2 i_3}
\cdots
\mathcal{O}_{i_n i_1}
}
&=
e^{-(n-1)S(E^+)}
F^{(n)}_{E^+}
(\omega_{i_1 i_2},\ldots,\omega_{i_{n-1}i_n}),
\end{align}
where $F^{(n)}_{E^+}(\vec{\omega})$ is a smooth function of the average energy
$E^+=(E_{i_1}+E_{i_2}+\cdots+E_{i_n})/n$ and of the $n-1$ independent energy differences collected in
$\vec{\omega}=(\omega_{i_1i_2},\ldots,\omega_{i_{n-1}i_n})$, where $\omega_{ij}=E_i-E_j$.

Additionally, when repeated indices appear in such a way that the cyclic product decomposes into disconnected loops, the statistical average factorizes in the thermodynamic limit. For example,
\begin{align}
\overline{
\mathcal{O}_{i_1 i_2}
\cdots
\mathcal{O}_{i_{k-1} i_1}
\mathcal{O}_{i_1 i_{k+1}}
\cdots
\mathcal{O}_{i_n i_1}
}
=
\overline{
\mathcal{O}_{i_1 i_2}
\cdots
\mathcal{O}_{i_{k-1} i_1}
}
\,
\overline{
\mathcal{O}_{i_1 i_{k+1}}
\cdots
\mathcal{O}_{i_n i_1}
}.
\end{align}
The full ETH ansatz can be motivated by a typicality argument based on the invariance of local observables under local rotations of the energy eigenbasis \cite{Foini:2018sdb,Pappalardi:2022aaz, Vallini:2025vvq}.

A remarkable consequence of the full ETH is that thermal free cumulants admit a particularly simple representation. The $n$-th thermal free cumulant is given solely by cyclic products involving distinct indices \cite{Pappalardi:2022aaz},
\begin{equation} \label{eq:ETHcumulant}
\kappa_n^{\rm ETH}\!\left(\mathcal{O}(t_1)\cdots \mathcal{O}(t_n)\right)
=
\frac{1}{Z}
\sum_{i_1\neq i_2\neq\cdots\neq i_n}
e^{-\beta E_{i_1}}
\mathcal{O}_{i_1i_2}\cdots \mathcal{O}_{i_ni_1}
e^{i(t_1\omega_{i_1i_2}+\cdots+t_n\omega_{i_ni_1})}.
\end{equation}
Here, the superscript ``ETH'' is used to emphasize that the above expression follows from the full ETH ansatz and need not hold in generic quantum systems.
Additionally, this expression can be written as \cite{Pappalardi:2022aaz}
\begin{equation} \label{eq:fullETHfrequency}
\kappa_n^{\rm ETH}
=
\mathrm{FT}\!\left[
F^{(n)}_{E_\beta}(\vec{\omega})
e^{-\beta\, \vec{\omega}\cdot\vec{\ell}_n}
\right]
\equiv
\int d\vec{\omega}\,
e^{i\vec{\omega}\cdot\vec{t}}\,
F^{(n)}_{E_\beta}(\vec{\omega})\,
e^{-\beta \, \vec{\omega}\cdot\vec{\ell}_n},
\end{equation}
where $\vec{\ell}_n=((n-1)/n,\ldots,1/n,0)$ encodes the thermal weighting of the energy differences, and $E_\beta=\langle H\rangle_\beta$ is the thermal expectation value of the total energy. This establishes a one-to-one correspondence between the hierarchy of ETH functions $F^{(n)}_{E_\beta}(\vec{\omega})$ and the hierarchy of thermal free cumulants $\kappa_n$.

Thus, the evaluation of generic correlation functions in systems satisfying the full ETH can be organized in two steps. First, one uses the moment--free-cumulant relation, Eq.~\eqref{eq:momentfromfreecumulants1}, to decompose the desired correlator into a sum of products of free cumulants. Then, each free cumulant appearing in this decomposition is evaluated using the ETH expression in Eq.~\eqref{eq:ETHcumulant}.

\subsection{Bounds on the large-frequency behavior of ETH smooth functions}

In this section, we discuss consistency conditions on the large-frequency behavior of the ETH smooth functions appearing in the full ETH ansatz. These conditions follow from the requirement that equal-time thermal free cumulants be finite. We first follow and extend the discussion of Ref.~\cite{Pappalardi:2022aaz}, which applies naturally to lattice models, and then discuss the case of continuum quantum field theories.
For notational simplicity, a free cumulant involving $n$ operators will sometimes be denoted as follows
\begin{equation}
    \kappa_n^{\rm ETH}(\mathcal{O}(t_1),\mathcal{O}(t_2),\cdots,\mathcal{O}(t_n))=\kappa_n^{\rm ETH}(t_1,t_2,\cdots,t_n)\,.
\end{equation}

\subsubsection{The case of lattice models}

We begin by discussing the dependence of the ETH smooth functions on the energy differences. To make the underlying structure transparent, we first consider the simpler case in which all insertions in the correlation function correspond to the same operator $\mathcal O$. For pedagogical purposes, we analyze the second-, third-, and fourth-order thermal free cumulants in detail. The general pattern will then become clear, allowing us to state the corresponding result for $n$-point-free cumulants.

\subsubsection*{Second-order thermal free-cumulant}
Assuming ETH, the second-order thermal free cumulant is given by
\begin{equation}
\kappa_2^{\rm ETH}\!\left(\mathcal O(t_1),\mathcal O(t_2)\right)
=
\frac{1}{Z}
\sum_{i_1\neq i_2}
e^{-\beta E_{i_1}}
\mathcal O_{i_1i_2}\mathcal O_{i_2i_1}
e^{i(\omega_1 t_1+\omega_2 t_2)} ,
\end{equation}
where
\begin{equation}
\omega_1=E_{i_1}-E_{i_2},
\qquad
\omega_2=E_{i_2}-E_{i_1}.
\end{equation}
The two frequencies are not independent, since
$\omega_1+\omega_2=0$. Moreover, by time-translation invariance, we may
set $t_2=0$ and define
\begin{equation}
t\equiv t_1-t_2,
\qquad
\omega\equiv\omega_1.
\end{equation}
The full-ETH ansatz then gives
\begin{equation}
\overline{
\mathcal O_{i_1i_2}\mathcal O_{i_2i_1}}
=
e^{-S(E^+)}F_{E^+}^{(2)}(\omega),
\qquad
E^+=\frac{E_{i_1}+E_{i_2}}{2}.
\end{equation}
Using a continuum approximation for the sums over energy eigenstates,
followed by a saddle-point approximation around the thermal energy
$E_\beta$, one obtains
\begin{equation}
\kappa_2^{\rm ETH}(\mathcal{O}(t),\mathcal{O}(0))
=
\int d\omega\,
F_{E_\beta}^{(2)}(\omega)
e^{i\omega t-\beta\omega/2}.
\label{eq:kappa2-fullETH}
\end{equation}

Notice that the orientation chosen for the index loop fixes the sign of
the frequency appearing in the ETH smooth function. Exchanging
$i_1\leftrightarrow i_2$ leaves $E^+$ unchanged but sends
$\omega\to-\omega$, so that
\begin{equation}
\overline{
\mathcal O_{i_2i_1}\mathcal O_{i_1i_2}}
=
e^{-S(E^+)}F_{E^+}^{(2)}(-\omega).
\end{equation}
Since the two insertions correspond to the same operator, the
matrix-element product is invariant under this exchange,
\begin{equation}
\mathcal O_{i_2i_1}\mathcal O_{i_1i_2}
=
\mathcal O_{i_1i_2}\mathcal O_{i_2i_1},
\end{equation}
and therefore
\begin{equation}
F_{E^+}^{(2)}(\omega)
=
F_{E^+}^{(2)}(-\omega).
\label{eq:F2-even}
\end{equation}

At equal times, Eq.~\eqref{eq:kappa2-fullETH} becomes
\begin{equation}
\kappa_2^{\rm ETH}(\mathcal{O}(0),\mathcal{O}(0))
=
\int_{-\infty}^{\infty}d\omega\,
F_{E_\beta}^{(2)}(\omega)e^{-\beta\omega/2}
=
2\int_0^\infty d\omega\,
F_{E_\beta}^{(2)}(\omega)
\cosh\left(\frac{\beta\omega}{2}\right),
\end{equation}
where in the second line we used
Eq.~\eqref{eq:F2-even}. Assuming that the equal-time cumulant is finite,
convergence at large frequency requires
\begin{equation}
\int_0^\infty d\omega\,
F_{E_\beta}^{(2)}(\omega)e^{\beta\omega/2}
<\infty.
\end{equation}
Thus, at the level of the exponential decay rate,
\begin{equation}
F_{E_\beta}^{(2)}(\omega)
\lesssim
e^{-\beta|\omega|/2},
\qquad
|\omega|\to\infty,
\end{equation}
up to subexponential factors compatible with convergence.

\subsubsection*{Third-order thermal free-cumulant}
Assuming ETH, the third-order thermal cumulant is given by:
\begin{equation} \label{eq:kappa3ETH}
    \kappa^{\rm ETH}_3( \mathcal{O}(t_1),\mathcal{O}(t_2),\mathcal{O}(t_3))
    =
    \frac{1}{Z}
    \sum_{i_1 \neq  i_2 \neq i_3}
    e^{-\beta E_{i_1}}
    e^{i \omega_1 t_1+i\omega_2 t_2+i \omega_3 t_3}
    \mathcal{O}_{i_1i_2}
    \mathcal{O}_{i_2i_3}
    \mathcal{O}_{i_3i_1},
\end{equation}
where
\begin{equation}
    \omega_1=E_{i_1}-E_{i_2},
    \qquad
    \omega_2=E_{i_2}-E_{i_3},
    \qquad
    \omega_3=E_{i_3}-E_{i_1}.
\end{equation}
Since these energy differences form a closed loop, they obey
\begin{equation}
    \omega_1+\omega_2+\omega_3=0,
\end{equation}
and therefore only two of them are independent, for instance
\begin{equation}
    \omega_3=-\omega_1-\omega_2.
\end{equation}
The full ETH ansatz for the cyclic product of matrix elements then takes the
form
\begin{equation}
    \overline{
    \mathcal{O}_{i_1i_2}
    \mathcal{O}_{i_2i_3}
    \mathcal{O}_{i_3i_1}
    }
    =
    e^{-2S(E^+)}
    F^{(3)}_{E^+}(\omega_1,\omega_2),
\end{equation}
where
\begin{equation}
    E^+=\frac{E_{i_1}+E_{i_2}+E_{i_3}}{3}.
\end{equation}
The smooth function $F^{(3)}_{E^+}$ has a cyclic symmetry. This is not an
additional dynamical assumption, but rather a consistency condition following
from the fact that the cyclic product of matrix elements has no preferred
starting point. Indeed, since the matrix elements are complex numbers, their
product is invariant under cyclic reordering:
\begin{equation}
    \mathcal{O}_{i_1i_2}
    \mathcal{O}_{i_2i_3}
    \mathcal{O}_{i_3i_1}
    =
    \mathcal{O}_{i_2i_3}
    \mathcal{O}_{i_3i_1}
    \mathcal{O}_{i_1i_2}.
\end{equation}
If we describe the same product starting at $i_2$, the corresponding
frequency variables are
\begin{equation}
    \omega_1'=E_{i_2}-E_{i_3}=\omega_2,
    \qquad
    \omega_2'=E_{i_3}-E_{i_1}=\omega_3.
\end{equation}
Therefore consistency of the ETH ansatz implies
\begin{equation}
    \overline{
    \mathcal{O}_{i_1i_2}
    \mathcal{O}_{i_2i_3}
    \mathcal{O}_{i_3i_1}
    }
    =
    \overline{
    \mathcal{O}_{i_2i_3}
    \mathcal{O}_{i_3i_1}
    \mathcal{O}_{i_1i_2}
    }
    =
    e^{-2S(E^+)}
    F^{(3)}_{E^+}(\omega_2,\omega_3).
\end{equation}
Repeating the same argument by starting the loop at $i_3$, one obtains
\begin{equation}
    \overline{
    \mathcal{O}_{i_1i_2}
    \mathcal{O}_{i_2i_3}
    \mathcal{O}_{i_3i_1}
    }
    =
    \overline{
    \mathcal{O}_{i_3i_1}
    \mathcal{O}_{i_1i_2}
    \mathcal{O}_{i_2i_3}
    }
    =
    e^{-2S(E^+)}
    F^{(3)}_{E^+}(\omega_3,\omega_1).
\end{equation}
Hence the smooth ETH function satisfies
\begin{equation} \label{eq:cyclicityETH}
    F^{(3)}_{E^+}(\omega_1,\omega_2)
    =
    F^{(3)}_{E^+}(\omega_2,\omega_3)
    =
    F^{(3)}_{E^+}(\omega_3,\omega_1),
    \qquad
    \omega_1+\omega_2+\omega_3=0.
\end{equation}

We now apply this condition to the third thermal free cumulant when full
ETH holds. Using Eq.~\eqref{eq:kappa3ETH}, we obtain
\begin{equation}
\begin{split}
\kappa_3^{\rm ETH}
\big(
\mathcal O(t_1),
\mathcal O(t_2),
\mathcal O(t_3)
\big)
=
\int d\omega_1\,d\omega_2\,
&F^{(3)}_{E_\beta}(\omega_1,\omega_2)
\exp\left[
-\frac{\beta}{3}(2\omega_1+\omega_2)
\right]
\\
&\times
e^{i\omega_1t_1+i\omega_2t_2+i\omega_3t_3},
\end{split}
\end{equation}
where $\omega_3=-\omega_1-\omega_2$. At equal times, the phase factor
drops out, and therefore
\begin{equation}
\kappa_3^{\rm ETH}
\big(
\mathcal O(0),\mathcal O(0),\mathcal O(0)
\big)
=
\int d\omega_1\,d\omega_2\,
F^{(3)}_{E_\beta}(\omega_1,\omega_2)
\exp\left[
-\frac{\beta}{3}(2\omega_1+\omega_2)
\right].
\label{eq:kappa3-equal-time-first}
\end{equation}

For a lattice system, the equal-time cumulant is expected to be finite.
Assuming absolute convergence, the smooth function must compensate any
exponential growth of the thermal factor. Equation
\eqref{eq:kappa3-equal-time-first} consequently gives
\begin{equation}
F^{(3)}_{E_\beta}(\omega_1,\omega_2)
\lesssim
\begin{cases}
\exp\left(-2\beta|\omega_1|/3\right),
& \omega_1\to-\infty,\quad \omega_2\ \text{fixed},\\[1mm]
\exp\left(-\beta|\omega_2|/3\right),
& \omega_2\to-\infty,\quad \omega_1\ \text{fixed}.
\end{cases}
\end{equation}

To constrain the positive-frequency limits, we cyclically change the
starting point of the index loop. Using the cyclicity property \eqref{eq:cyclicityETH},
the equal-time cumulant admits the equivalent representations
\begin{equation}
\kappa_3^{\rm ETH}
=
\int d\omega_1\,d\omega_2\,
F^{(3)}_{E_\beta}(\omega_1,\omega_2)
\exp\left[
-\frac{\beta}{3}(-\omega_1+\omega_2)
\right],
\label{eq:kappa3-equal-time-second}
\end{equation}
and
\begin{equation}
\kappa_3^{\rm ETH}
=
\int d\omega_1\,d\omega_2\,
F^{(3)}_{E_\beta}(\omega_1,\omega_2)
\exp\left[
-\frac{\beta}{3}(-\omega_1-2\omega_2)
\right].
\label{eq:kappa3-equal-time-third}
\end{equation}
Equation~\eqref{eq:kappa3-equal-time-second} constrains
$\omega_1\to+\infty$, while Eq.~\eqref{eq:kappa3-equal-time-third}
constrains $\omega_2\to+\infty$. Combining all four coordinate limits,
we find
\begin{equation}
F^{(3)}_{E_\beta}(\omega_1,\omega_2)
\lesssim
\begin{cases}
\exp\left(-\beta\omega_1/3\right),
& \omega_1\to+\infty,\\[1mm]
\exp\left(-2\beta|\omega_1|/3\right),
& \omega_1\to-\infty,
\end{cases}
\qquad \omega_2\ \text{fixed},
\end{equation}
and
\begin{equation}
F^{(3)}_{E_\beta}(\omega_1,\omega_2)
\lesssim
\begin{cases}
\exp\left(-2\beta\omega_2/3\right),
& \omega_2\to+\infty,\\[1mm]
\exp\left(-\beta|\omega_2|/3\right),
& \omega_2\to-\infty,
\end{cases}
\qquad \omega_1\ \text{fixed}.
\end{equation}

\subsubsection*{Fourth-order thermal free cumulant}

When full ETH holds, the fourth-order thermal free cumulant is given by
\begin{equation}
\label{eq:kappa4ETH}
\begin{split}
\kappa_4^{\rm ETH}(t_1,t_2,t_3,t_4)
\equiv{}&
\kappa_4^{\rm ETH}
\big(
\mathcal O(t_1),
\mathcal O(t_2),
\mathcal O(t_3),
\mathcal O(t_4)
\big)
\\
={}&
\frac{1}{Z}
\sum_{i_1\neq i_2 \neq i_3 \neq i_4}
e^{-\beta E_{i_1}}
\mathcal O_{i_1i_2}
\mathcal O_{i_2i_3}
\mathcal O_{i_3i_4}
\mathcal O_{i_4i_1}
e^{i\sum_{a=1}^{4}\omega_a t_a}.
\end{split}
\end{equation}
The frequencies are
\begin{equation}
\omega_1=E_{i_1}-E_{i_2},
\qquad
\omega_2=E_{i_2}-E_{i_3},
\qquad
\omega_3=E_{i_3}-E_{i_4},
\qquad
\omega_4=E_{i_4}-E_{i_1},
\end{equation}
and satisfy
\begin{equation}
\omega_1+\omega_2+\omega_3+\omega_4=0.
\end{equation}
The full-ETH ansatz reads
\begin{equation}
\overline{
\mathcal O_{i_1i_2}
\mathcal O_{i_2i_3}
\mathcal O_{i_3i_4}
\mathcal O_{i_4i_1}}
=
e^{-3S(E^+)}
F_{E^+}^{(4)}(\omega_1,\omega_2,\omega_3),
\qquad
E^+=\frac{1}{4}\sum_{a=1}^{4}E_{i_a}.
\end{equation}
After the continuum and saddle-point approximations, one obtains
\begin{equation}
\kappa_4^{\rm ETH}(t_1,t_2,t_3,t_4)
=
\int d\omega_1\,d\omega_2\,d\omega_3\,
F_{E_\beta}^{(4)}(\omega_1,\omega_2,\omega_3)
e^{
-\frac{\beta}{4}
\left(3\omega_1+2\omega_2+\omega_3\right)
}
e^{i\sum_{a=1}^{4}\omega_a t_a},
\label{eq:kappa4-fourier}
\end{equation}
where $\omega_4=-\omega_1-\omega_2-\omega_3$.
The cyclicity property implies
\begin{equation}
F^{(4)}(\omega_1,\omega_2,\omega_3)
=
F^{(4)}(\omega_2,\omega_3,\omega_4)
=
F^{(4)}(\omega_3,\omega_4,\omega_1)
=
F^{(4)}(\omega_4,\omega_1,\omega_2).
\end{equation}
Consequently, at equal times, the free cumulant admits four equivalent
representations with thermal factors
\begin{equation}
\begin{aligned}
\mathcal L_1&=3\omega_1+2\omega_2+\omega_3,\\
\mathcal L_2&=-\omega_1+2\omega_2+\omega_3,\\
\mathcal L_3&=-\omega_1-2\omega_2+\omega_3,\\
\mathcal L_4&=-\omega_1-2\omega_2-3\omega_3,
\end{aligned}
\qquad
\kappa_4^\beta
=
\int d^3\omega\,
F_{E_\beta}^{(4)}(\vec\omega)
e^{-\beta\mathcal L_a/4}.
\end{equation}

Assuming absolute convergence, these representations imply the
following bounds
\begin{equation}
F_{E_\beta}^{(4)}(\omega_1,\omega_2,\omega_3)
\lesssim
\begin{cases}
e^{-\beta\omega_1/4},
& \omega_1\to+\infty,\\
e^{-3\beta|\omega_1|/4},
& \omega_1\to-\infty,
\end{cases}
\qquad
\omega_2,\omega_3\ \mathrm{fixed},
\end{equation}
\begin{equation}
F_{E_\beta}^{(4)}(\omega_1,\omega_2,\omega_3)
\lesssim
e^{-\beta|\omega_2|/2},
\qquad
|\omega_2|\to\infty,
\qquad
\omega_1,\omega_3\ \mathrm{fixed},
\end{equation}
and
\begin{equation}
F_{E_\beta}^{(4)}(\omega_1,\omega_2,\omega_3)
\lesssim
\begin{cases}
e^{-3\beta\omega_3/4},
& \omega_3\to+\infty,\\
e^{-\beta|\omega_3|/4},
& \omega_3\to-\infty,
\end{cases}
\qquad
\omega_1,\omega_2\ \mathrm{fixed}.
\end{equation}

\subsubsection*{$n$th-order thermal free cumulant}

Proceeding as in the previous examples, one finds that, for an
$n$-point free cumulant, the large-frequency bounds take
the form
\begin{equation}
F^{(n)}_{E_\beta}(\omega_1,\ldots,\omega_{n-1})
\lesssim
\begin{cases}
\displaystyle
\exp\left(-\frac{k\beta}{n}\,\omega_k\right),
& \omega_k\to+\infty,\\[2mm]
\displaystyle
\exp\left(-\frac{(n-k)\beta}{n}\,|\omega_k|\right),
& \omega_k\to-\infty,
\end{cases}
\qquad
k=1,\ldots,n-1,
\label{eq:general-coordinate-bound}
\end{equation}
where all the other independent frequencies are kept fixed and
$\omega_n=-\sum_{j=1}^{n-1}\omega_j$. As before, these expressions
specify the exponential decay rates, up to subexponential factors
compatible with convergence.

It is worth emphasizing that the bounds in
Eq.~\eqref{eq:general-coordinate-bound} are generally asymmetric under
$\omega_k\to-\omega_k$ and depend on which independent frequency is
taken to be large. This reflects the choice of frequency coordinates in
which the $n$th insertion is fixed at $t_n=0$, so that different
coordinate directions correspond to inequivalent limits in the
constrained frequency space. In the next section, we will see that, in
QFT, these decay rates admit a simple geometrical interpretation in
terms of the distances of the operator insertions from the collision
boundaries on the thermal circle. The largest coordinate-wise decay rate is
$(n-1)\beta/n$, obtained, for example, when
$\omega_1\to-\infty$ or equivalently when
$\omega_{n-1}\to+\infty$. This agrees with the extremal scale identified
in Ref.~\cite{Pappalardi:2022aaz}. The opposite directions and the
intermediate frequencies probe different portions of the thermal
circle and are consequently governed by different exponential rates.

\subsubsection*{Additional symmetry $F_{E^+}^{(n)}(\vec{\omega})
=F_{E^+}^{(n)}(-\vec{\omega})$}

If the full-ETH smooth function possesses the additional inversion symmetry
\begin{equation}
F_{E^+}^{(n)}(\vec{\omega})
=
F_{E^+}^{(n)}(-\vec{\omega})\,,
\end{equation}
the decay along opposite frequency directions must be identical. Consequently,
the bounds obtained for $\omega_k\rightarrow+\infty$ and
$\omega_k\rightarrow-\infty$ must hold simultaneously, and the stronger of
the two determines the decay in both directions.

\subsubsection*{A subtlety with different operator insertions}

When the insertions are not all associated with the same operator, as in
\begin{equation}
\kappa_n^{\rm ETH}
\big(
\mathcal O_1(t_1),\ldots,\mathcal O_n(t_n)
\big),
\end{equation}
their labels must be retained explicitly in the corresponding full-ETH
smooth function,
\begin{equation}
F^{(n)}_{E^+;\,\mathcal O_1\cdots\mathcal O_n}
(\omega_1,\ldots,\omega_{n-1}) .
\end{equation}
Cyclicity then relates smooth functions with cyclically permuted operator
orderings,
\begin{equation}
F^{(n)}_{E^+;\,\mathcal O_1\cdots\mathcal O_n}
(\omega_1,\ldots,\omega_{n-1})
=
F^{(n)}_{E^+;\,\mathcal O_2\cdots\mathcal O_n\mathcal O_1}
(\omega_2,\ldots,\omega_n),
\end{equation}
where
\begin{equation}
\omega_n=-\sum_{a=1}^{n-1}\omega_a.
\end{equation}
Thus, unlike the case of identical insertions, cyclicity does not
necessarily relate the same smooth function evaluated at different
frequency arguments.

A relevant example is the holographic out-of-time-order correlator
\begin{equation*}
\left\langle
V(t_1)W(t_2)V(t_3)W(t_4)
\right\rangle_\beta ,
\end{equation*}
whose irreducible contribution is described by
$F^{(4)}_{E^+;\,VWVW}$. A cyclic shift by one insertion relates it to
$F^{(4)}_{E^+;\,WVWV}$, while a shift by two insertions returns to the
original operator ordering. 

\subsubsection{The case of continuum quantum field theories}

The derivation of Eq.~\eqref{eq:general-coordinate-bound} given above
relies on the finiteness of equal-time thermal free cumulants. While this
is natural in lattice systems with bounded local operators, it does not
directly extend to continuum quantum field theories. Indeed, when local
operators become coincident, their correlation functions generally
develop ultraviolet singularities, so the corresponding equal-time free
cumulants need not be finite. We therefore regulate the cumulants by
placing the operator insertions at equal intervals along the Euclidean
thermal circle. The frequency-space bounds can then be derived from the
analyticity of these regulated correlators in complex time, together with
the assumption that their singularities near the collision boundaries
are at most power-law.

Consider the $n$-point regulated free
cumulant:
\begin{equation}
\kappa_n^{\rm reg}(\vec t)
=
\kappa_n^{\rm reg} (
y_n\mathcal O_1(t_1),y_n\mathcal O_2(t_2),
\cdots,
y_n\mathcal O_n(t_n)).
\end{equation}
where
\begin{equation}
y_n=\frac{e^{-\beta H/n}}{Z^{1/n}}\,, \quad \quad \vec t =(t_1,t_2,\cdots,t_n)\,.
\end{equation}
Under full ETH, these cumulants are given by
the cyclic contribution with all energy indices distinct,
\begin{equation}
\begin{split}
\kappa_n^{\rm ETH,reg}(\vec t)
=
\frac{1}{Z}
\sum_{i_1 \neq i_2 \neq  \ldots \neq i_n}
&
e^{
-\frac{\beta}{n}\sum_{a=1}^{n}E_{i_a}}
 O_{i_1 i_2} O_{i_2 i_3} \cdots  O_{i_n i_1}
e^{i\sum_{a=1}^{n}\omega_at_a},
\end{split}
\end{equation}
where
\begin{equation}
\omega_a=E_{i_a}-E_{i_{a+1}},
\qquad
\sum_{a=1}^{n}\omega_a=0.
\end{equation}
Since
\begin{equation}
E^+=\frac{1}{n}\sum_{a=1}^{n}E_{i_a},
\end{equation}
the symmetric regulator contributes simply $e^{-\beta E^+}$. Applying
the full-ETH ansatz and performing the thermal saddle then gives
\begin{equation}
\kappa_n^{\rm ETH,reg}(\vec t)
=
\int d^{\,n-1}\omega\,
F^{(n)}_{E_\beta}(\omega_1,\ldots,\omega_{n-1})
e^{i\sum_{a=1}^{n}\omega_at_a}.
\end{equation}
Using time-translation invariance, we set $t_n=0$ and shift the remaining
contours according to
\begin{equation}
t_a\longrightarrow t_a+i\eta_a,
\qquad
a=1,\ldots,n-1.
\end{equation}
The Euclidean gaps between neighboring operator insertions are
\begin{equation}
a_1=\frac{\beta}{n}+\eta_1,
\qquad
a_j=\frac{\beta}{n}-\eta_{j-1}+\eta_j,
\quad j=2,\ldots,n-1,
\end{equation}
and
\begin{equation}
a_n=\frac{\beta}{n}-\eta_{n-1}.
\end{equation}
The common analyticity domain of the regulated moments, and hence of
their free-cumulant combination, is
\begin{equation}
D_n=
\left\{
\vec\eta\in\mathbb R^{n-1}:a_j>0,\quad j=1,\ldots,n
\right\}.
\end{equation}
For any fixed $\vec\eta\in\mathcal D_n$, analyticity allows the integration contours to be shifted without crossing singularities, giving
\begin{equation}
F^{(n)}_{E_\beta}(\vec\omega)
=
\int\frac{d^{n-1}t}{(2\pi)^{n-1}}\,
e^{-i\vec\omega\cdot(\vec t+i\vec\eta)}
\kappa_n^{\mathrm{reg}}(\vec t+i\vec\eta)\,.
\label{eq:deformed-Fourier}
\end{equation}
Taking the absolute value then yields
\begin{equation}
\left|
F^{(n)}_{E_\beta}(\vec\omega)
\right|
\leq
C(\vec\eta)\,
e^{\vec\omega\cdot\vec\eta},
\qquad
\vec\eta\in D_n,
\end{equation}
where
\begin{equation}
C(\vec\eta)
=
\int\frac{d^{\,n-1}t}{(2\pi)^{n-1}}
\left|
\kappa_n^{\rm ETH,reg}(\vec t+i\vec\eta)
\right|.
\end{equation}
Here we make an explicit boundedness assumption that for contours in the interior of $\mathcal D_n$, $C(\vec\eta)$ is finite, and as the contour approaches a boundary corresponding to an operator collision, it grows at most algebraically in the inverse Euclidean separation. In a local QFT this behavior is naturally suggested by the operator-product expansion, while sufficient real-time clustering is assumed to control the integral away from the collision regions. In particular, $C(\vec\eta)$ must not generate an additional exponential dependence on the large-frequency scale considered below. This assumption plays a role analogous to the boundedness condition entering the derivation of the conventional chaos bound in \cite{Maldacena:2015waa}.

Let us first examine explicitly the direction associated with
$\omega_1$. From the positivity of the first Euclidean gap,
\begin{equation}
a_1=\frac{\beta}{n}+\eta_1>0,
\end{equation}
we immediately obtain
\begin{equation}
\eta_1>-\frac{\beta}{n}.
\end{equation}
Geometrically, the lower endpoint corresponds to moving the first
insertion, located at $t_1+i\eta_1$, toward the fixed insertion at
$t_n=0$ across the adjacent thermal gap. Since the initial separation
between them is $\beta/n$, the first operator can be displaced by at
most $\beta/n$ in this direction before the two insertions collide.

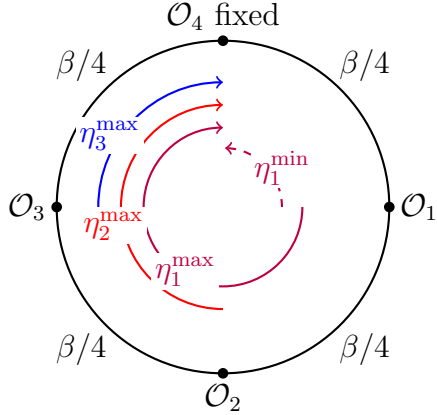
\begin{figure}[t]
\centering
\begin{tikzpicture}[scale=1.0]

% Thermal circle
\draw[thick] (0,0) circle (2.2);

% Initial equally spaced positions
\coordinate (O4) at (90:2.2);
\coordinate (O3) at (180:2.2);
\coordinate (O2) at (270:2.2);
\coordinate (O1) at (0:2.2);

% Points
\fill (O4) circle (2pt);
\fill (O3) circle (2pt);
\fill (O2) circle (2pt);
\fill (O1) circle (2pt);

% Labels
\node[above] at (O4) {$\mathcal O_4$ fixed};
\node[left] at (O3) {$\mathcal O_3$};
\node[below] at (O2) {$\mathcal O_2$};
\node[right] at (O1) {$\mathcal O_1$};

% Gap labels
\node at (135:2.65) {$\beta/4$};
\node at (225:2.65) {$\beta/4$};
\node at (315:2.65) {$\beta/4$};
\node at (45:2.65) {$\beta/4$};

% Arrows showing possible shifts toward O4
\draw[->, thick, blue]
  (180:1.65) arc[start angle=180,end angle=90,radius=1.65];

\draw[->, thick, red]
  (270:1.35) arc[start angle=270,end angle=90,radius=1.35];

\draw[->, thick, purple]
  (0:1.05) arc[start angle=0,end angle=-270,radius=1.05];

% Eta labels placed freely, independent of the arcs
\node[blue, fill=white, inner sep=1.5pt] at (-1.50,0.95) {$\eta_3^{\max}$};
\node[red, fill=white, inner sep=1.5pt] at (-1.45,-0.25) {$\eta_2^{\max}$};
\node[purple, fill=white, inner sep=1.5pt] at (-0.5,-0.85) {$\eta_1^{\max}$};

% Center label
%\node at (0,0) {thermal circle};

% Dashed arrow for eta_1^min = -beta/4
\draw[->, thick, dashed, purple]
  (0:0.78) arc[start angle=0,end angle=88,radius=0.78];

  \node[purple, fill=white, inner sep=1.2pt] at (0.78,0.52) {$\eta_1^{\min}$};

\end{tikzpicture}
\caption{ \justifying
Imaginary-time deformations of four equally spaced operator
insertions on the thermal circle. Fixing $\mathcal O_4$ by
time-translation invariance, the boundary of the analyticity domain
allows maximal positive displacements
$\eta_3^{\max}=\beta/4$,
$\eta_2^{\max}=\beta/2$, and
$\eta_1^{\max}=3\beta/4$ while preserving the cyclic ordering.
For $\mathcal O_1$, the opposite boundary is
$\eta_1^{\min}=-\beta/4$ (dashed line). Therefore 
$-\beta/4<\eta_1<3\beta/4$.
}
\label{fig:kms-four-point-shift}
\end{figure}

The opposite endpoint follows from the remaining gaps. Their sum is
\begin{equation}
\sum_{j=2}^{n}a_j
=
\frac{(n-1)\beta}{n}-\eta_1.
\end{equation}
Since each gap must remain positive, their sum must also be positive,
which gives
\begin{equation}
\eta_1<\frac{(n-1)\beta}{n}.
\end{equation}
Geometrically, this corresponds to moving the first insertion in the
opposite direction around the thermal circle, while allowing the
intermediate insertions to move so that their cyclic ordering is
preserved. The first operator can then traverse the remaining $n-1$
thermal gaps, for a total imaginary-time displacement approaching
$(n-1)\beta/n$, before colliding with the fixed insertion from the
opposite side. Figure~\ref{fig:kms-four-point-shift} illustrates this
argument for the $n=4$ case. The projection of the analyticity domain
onto the $\eta_1$ axis is therefore
\begin{equation}
-\frac{\beta}{n}
<
\eta_1
<
\frac{(n-1)\beta}{n}.
\label{eq:eta1-projection}
\end{equation}

Consider first $\omega_1=W\to+\infty$, with the remaining independent
frequencies held fixed. The contour deformation produces the factor $e^{\omega_1\eta_1}=e^{W\eta_1}.$ The strongest suppression is obtained by approaching the lower endpoint
of Eq.~\eqref{eq:eta1-projection},
\begin{equation}
\eta_1\to-\frac{\beta}{n},
\end{equation}
which yields
\begin{equation}
\left|
F^{(n)}_{E_\beta}
(W,\omega_2,\ldots,\omega_{n-1})
\right|
\lesssim
\operatorname{poly}(W)
\exp\left(-\frac{\beta}{n}W\right).
\label{eq:positive-omega1-bound}
\end{equation}

For the opposite direction, $\omega_1=-W\to-\infty$, the contour factor
takes the following form $e^{\omega_1\eta_1}=e^{-W\eta_1}.$ The strongest suppression is now obtained by approaching the upper
endpoint,
\begin{equation}
\eta_1\to\frac{(n-1)\beta}{n},
\end{equation}
and hence
\begin{equation}
\left|
F^{(n)}_{E_\beta}
(-W,\omega_2,\ldots,\omega_{n-1})
\right|
\lesssim
\operatorname{poly}(W)
\exp\left[-\frac{(n-1)\beta}{n}W\right].
\label{eq:negative-omega1-bound}
\end{equation}
Thus, for $n>2$, the two directions along the $\omega_1$ axis are
generally inequivalent. 

More generally, the allowed range of $\eta_k$ follows directly by
summing the gaps on the two sides of the $k$th insertion. The sum of the
first $k$ gaps is
\begin{equation}
\sum_{j=1}^{k}a_j
=
\frac{k\beta}{n}+\eta_k,
\end{equation}
while the sum of the remaining gaps is
\begin{equation}
\sum_{j=k+1}^{n}a_j
=
\frac{(n-k)\beta}{n}-\eta_k.
\end{equation}
Positivity of these two sums gives
\begin{equation}
-\frac{k\beta}{n}
<
\eta_k
<
\frac{(n-k)\beta}{n},
\qquad
k=1,\ldots,n-1.
\label{eq:etak-projection}
\end{equation}

For $\omega_k=W\to+\infty$, the contour factor is $e^{W\eta_k}$, so
the optimal contour approaches the lower endpoint of
Eq.~\eqref{eq:etak-projection}. For $\omega_k=-W\to-\infty$, the
factor is $e^{-W\eta_k}$, and the optimal contour approaches the upper
endpoint. Therefore, for each $k=1,\ldots,n-1$, one finds
\begin{equation}
\left|
F^{(n)}_{E_\beta}(\omega_1,\ldots,\omega_{n-1})
\right|
\lesssim
\begin{cases}
\operatorname{poly}(\omega_k)
\exp\left(-\dfrac{k\beta}{n}\omega_k\right),
& \omega_k\to+\infty,\\[2mm]
\operatorname{poly}(|\omega_k|)
\exp\left(-\dfrac{(n-k)\beta}{n}|\omega_k|\right),
& \omega_k\to-\infty,
\end{cases}
\label{eq:QFT-general-coordinate-bound}
\end{equation}
with all other independent frequencies held fixed. These bounds coincide with those derived for lattice models in
Eq.~\eqref{eq:general-coordinate-bound}. In the QFT setting, however,
their geometrical origin becomes more transparent: they follow directly
from the analyticity domain of regulated free cumulants in complex time. For reference, the first few cases are summarized in
Table~\ref{tab:eth-frequency-bounds}. Defining the decay coefficients by
\begin{equation}
|F_{E_{\beta}}^{(n)}|
\lesssim
\operatorname{poly}(|\omega_k|)
e^{-\beta\alpha_{n,k}^{\pm}|\omega_k|},
\qquad
\omega_k\to\pm\infty,
\end{equation}
Eq.~\eqref{eq:QFT-general-coordinate-bound} gives
\begin{equation}
\alpha_{n,k}^{+}=\frac{k}{n},
\qquad
\alpha_{n,k}^{-}=\frac{n-k}{n}.
\end{equation}

\begin{table}[h]
\centering
\begin{tabular}{c|ccc}
\hline\hline
$n$ &
$k=1$ &
$k=2$ &
$k=3$
\\
\hline
$2$ &
$\left(\frac12,\frac12\right)$ &
-- &
--
\\[2mm]
$3$ &
$\left(\frac13,\frac23\right)$ &
$\left(\frac23,\frac13\right)$ &
--
\\[2mm]
$4$ &
$\left(\frac14,\frac34\right)$ &
$\left(\frac12,\frac12\right)$ &
$\left(\frac34,\frac14\right)$
\\
\hline\hline
\end{tabular}
\caption{
Large-frequency decay coefficients
$(\alpha_{n,k}^{+},\alpha_{n,k}^{-})$ for the
$n$th-order full-ETH smooth function. The first (second) entry in each
pair corresponds to $\omega_k\to+\infty$
($\omega_k\to-\infty$), with all other independent frequencies held
fixed.
}
\label{tab:eth-frequency-bounds}
\end{table}

\newpage

In the next section, we show that two-dimensional CFTs saturate the
bounds for second- and third-order thermal free cumulants, whose forms are
fixed by conformal symmetry up to standard CFT data. The first genuinely
dynamical test arises at fourth order, where the correlator depends
nontrivially on conformal cross-ratios and saturation is therefore not
expected for a generic CFT. We nevertheless show that, for holographic
CFTs, the strongest fourth-order bound is saturated by the leading
gravitational contribution, computed in the bulk eikonal approximation
in the Regge regime.

\section{Saturation of the bounds in thermal CFTs}
\label{sec:CFTs}

In this section, we test the frequency-space bounds
\eqref{eq:QFT-general-coordinate-bound} in thermal conformal field theories,
without assuming a priori that full ETH holds. We begin with the universal
two- and three-point functions of CFT$_2$, whose forms are fixed by conformal
symmetry, and compare the large-frequency behavior of their regulated free
cumulants with the corresponding full-ETH bounds. We then turn to the
fourth-order free cumulant, where genuine dynamical information first appears,
and analyze its leading holographic contribution in the gravitational Regge
regime.

\subsection{Two- and three-point functions in thermal CFT$_2$}

We first consider scalar primary operators inserted at the same spatial point.
Their finite-temperature two- and three-point functions on the infinite line
follow from the standard conformal map from the complex plane to the thermal
cylinder,
\begin{equation}
z=e^{\frac{2\pi}{\beta}w}~,
\qquad
w=x+i\tau~,
\qquad {\rm and} \quad
\tau\sim\tau+\beta~.
\label{eq:plane-cylinder-map}
\end{equation}
Setting the spatial separations to zero, $x_i=0$, the thermal two- and three-point functions take the universal forms
\begin{align}
\left\langle
\mathcal O(\tau_1)
\mathcal O(\tau_2)
\right\rangle_\beta
&=
\left(\frac{\pi}{\beta}\right)^{2\Delta}
\frac{c_{\mathcal O}}{
\left|
\sin\left(\frac{\pi\tau_{12}}{\beta}\right)
\right|^{2\Delta}}
\,,
\\
\left\langle
\mathcal O_1(\tau_1)
\mathcal O_2(\tau_2)
\mathcal O_3(\tau_3)
\right\rangle_\beta
&=
\left(\frac{\pi}{\beta}\right)^{\Delta_1+\Delta_2+\Delta_3}
\nonumber 
\quad
\frac{c_{123}}{
\left|
\sin\left(\frac{\pi\tau_{12}}{\beta}\right)
\right|^{\delta_{12}}
\left|
\sin\left(\frac{\pi\tau_{23}}{\beta}\right)
\right|^{\delta_{23}}
\left|
\sin\left(\frac{\pi\tau_{31}}{\beta}\right)
\right|^{\delta_{31}}
}
\,,
\end{align}
where $\tau_{ij}=\tau_i-\tau_j$ and
\begin{equation}
\delta_{12}=\Delta_1+\Delta_2-\Delta_3~,
\qquad
\delta_{23}=\Delta_2+\Delta_3-\Delta_1~,
\qquad
\delta_{31}=\Delta_3+\Delta_1-\Delta_2\,.
\label{eq:delta-definitions}
\end{equation}
These expressions are completely fixed by conformal symmetry up to the operator normalization $c_{\mathcal O}$ and the three-point coefficient $c_{123}$.

We consider centered operators, such that their thermal one-point functions vanish and the second- and third-order free cumulants coincide with the corresponding correlation functions. We further consider equally spaced operator insertions along the thermal circle. For the two-point function, the two operators are separated by $\beta/2$, and the regulated thermal free cumulant is
\begin{equation}
\kappa^{\rm reg}_2(t)
=\left\langle
\mathcal O\left(t-\frac{i\beta}{2}\right)
\mathcal O(0)
\right\rangle_\beta
=
c_{\mathcal O}
\left(\frac{\pi}{\beta}\right)^{2\Delta}
\frac{1}{
\cosh^{2\Delta}\left(\frac{\pi t}{\beta}\right)}
\,.
\label{eq:CFT-kappa2-reg}
\end{equation}
It is convenient to introduce the shifted thermal kernel
\begin{equation}
K_{\delta,a}(t)
\equiv
\left[
i\sinh\left(
\frac{\pi}{\beta}(t-ia)
\right)
\right]^{-\delta}\,.
\label{eq:CFT-shifted-kernel}
\end{equation}
In terms of this kernel, the three-point function with equally spaced operators on the thermal circle can be written as
\begin{align}
\kappa^{\rm reg}_3(t_1,t_2)
&=
\left\langle
\mathcal O_1\left(t_1-\frac{2i\beta}{3}\right)
\mathcal O_2\left(t_2-\frac{i\beta}{3}\right)
\mathcal O_3(0)
\right\rangle_\beta
\nonumber\\
&=
c_{123}
\left(\frac{\pi}{\beta}\right)^{\Delta_1+\Delta_2+\Delta_3}
K_{\delta_{12},\beta/3}(t_1-t_2)
K_{\delta_{23},\beta/3}(t_2)
K_{\delta_{31},2\beta/3}(t_1)\,.
\label{eq:CFT-kappa3-reg}
\end{align}

We now proceed to compute the multitime Fourier transforms of the above free cumulants and show that they saturate the full-ETH bounds. We begin with the Fourier transform of the shifted kernel,
\begin{equation} 
\widetilde K_{\delta,a}(\omega)
=\int_{-\infty}^{\infty}dt\,
e^{-i\omega t}
K_{\delta,a}(t)\,.
\label{eq:CFT-kernel-FT-def}
\end{equation}
Using the integral representation of the Beta function one can show that
\begin{equation} \label{eq:FourierKernel}
\widetilde K_{\delta,a}(\omega)
=e^{-\left(\frac{\beta}{2}-a\right)\omega}
\frac{\beta}{\pi}\,
\frac{2^{\delta-1}}{\Gamma(\delta)}
\Gamma\left(
\frac{\delta}{2}+\frac{i\beta\omega}{2\pi}
\right)
\Gamma\left(
\frac{\delta}{2}-\frac{i\beta\omega}{2\pi}
\right)\,.
\end{equation}
Using Stirling's formula, one can see that the large frequency behavior is given by
\begin{equation}
\widehat{K}_{\delta,a}(\omega)
\sim
\operatorname{poly}(|\omega|)
\exp\left[
\left(a-\frac{\beta}{2}\right)\omega
-\frac{\beta}{2}|\omega|
\right]\,.
\label{eq:K-asymptotic-general}
\end{equation}
It is useful to define the positive rate function
\begin{equation}
r_a(\omega)
\equiv
\frac{\beta}{2}|\omega|
-
\left(a-\frac{\beta}{2}\right)\omega
=
\begin{cases}
(\beta-a)\omega\,,
&\omega>0\,,
\\[2mm]
a|\omega|\,,
& \omega<0\,.
\end{cases}
\label{eq:rate-function}
\end{equation}
Then
\begin{equation}
\widehat{K}_{\delta,a}(\omega)
\sim
\operatorname{poly}(|\omega|)
e^{-r_a(\omega)}= \operatorname{poly}(|\omega|)
\begin{cases}
e^{-(\beta-a)\omega}\,,
& \omega\rightarrow+\infty\,,
\\[2mm]
e^{-a|\omega|}\,,
& \omega\rightarrow-\infty\,.
\end{cases}
\label{eq:K-asymptotic-directions}
\end{equation}
Note that the shifted kernel corresponds to a thermal two-point function with a generic imaginary-time separation $K_{\delta,a}(t)
\propto
\left\langle
\mathcal O(t-ia)\mathcal O(0)
\right\rangle_\beta
$ for an operator of scaling dimension $\Delta=\delta/2$. Its singularities occur whenever $t=i(a+n\beta)$ with $n\in\mathbb Z$. For $0<a<\beta$, the singularities closest to the real axis are therefore located at $t=ia$ in the upper half-plane and at $t=-i(\beta-a)$ in the lower half-plane.  
The asymmetric large-frequency decay \eqref{eq:K-asymptotic-directions} is therefore a direct consequence of the asymmetric location of the nearest thermal singularities relative to the real-time contour.

\paragraph{Second free-cumulant.}
For the regulated two-point cumulant, $a=\beta/2$ and $\delta=2\Delta$, so that the exponential prefactor in Eq.~\eqref{eq:FourierKernel} is absent. We therefore find
\begin{equation}
\widetilde\kappa^{\rm reg}_2(\omega)
=
c_{\mathcal O}
\left(\frac{\pi}{\beta}\right)^{2\Delta}
\frac{\beta}{\pi}\,
\frac{2^{2\Delta-1}}{\Gamma(2\Delta)}
\left|
\Gamma\left(
\Delta+\frac{i\beta\omega}{2\pi}
\right)
\right|^2\,.
\label{eq:CFT-kappa2-FT}
\end{equation}
At large frequencies we find
\begin{equation}
\widetilde\kappa^{\rm reg}_2(\omega)
\sim
\operatorname{poly}(|\omega|)
e^{-\frac{\beta}{2}|\omega|}\,,
\qquad
|\omega|\rightarrow\infty\,.
\label{eq:CFT-kappa2-asymptotic}
\end{equation}
This behavior could also have been anticipated directly from the analytic structure of the real-time correlator. The
singularities of $\cosh^{2\Delta}\left(\frac{\pi t}{\beta}\right)$ 
nearest to the real-time axis are located symmetrically at $t=\pm i\beta/2$. Their distance $\beta/2$ from the real axis determines the exponential decay $e^{-\beta|\omega|/2}$ of the Fourier transform. Thus, the Fourier transform of the regulated second free cumulant realizes precisely the minimal exponential decay allowed by the second-order full-ETH bound~\eqref{eq:QFT-general-coordinate-bound}.

\begin{figure}
    \centering
    \includegraphics[width=1\linewidth]{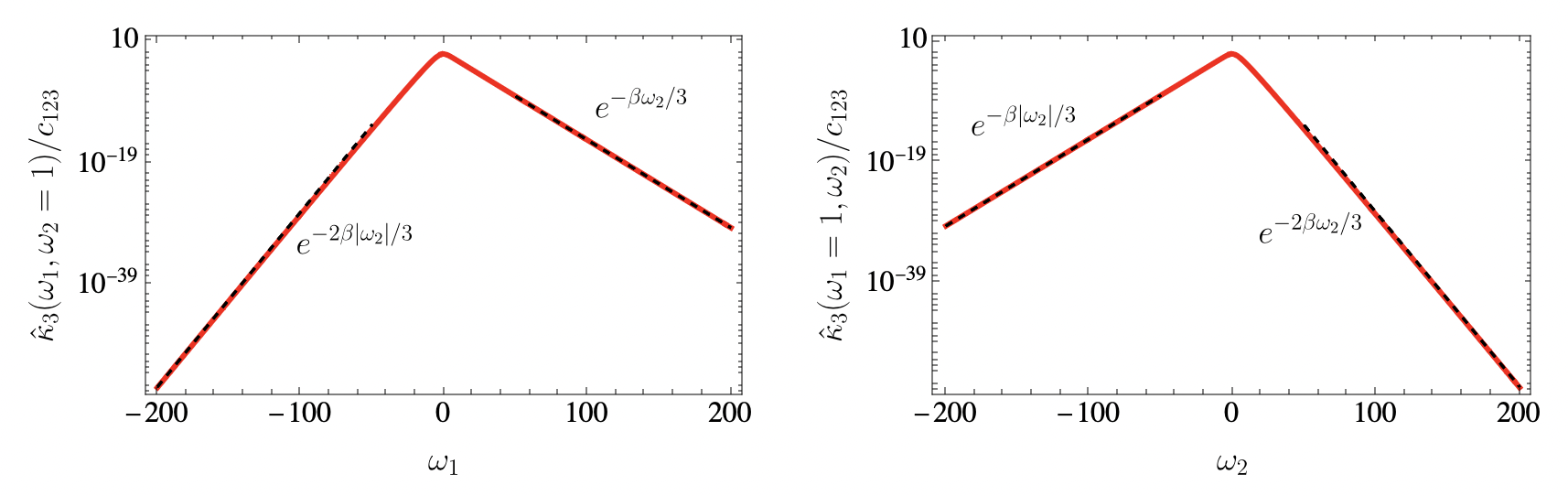}
    \caption{Large-frequency behavior of the Fourier transform of the third regulated free cumulant $\widehat{\kappa}_3(\omega_1,\omega_2)$. Here we set $\beta=1$ and $\Delta_1=\Delta_2=\Delta_3=1$. In the left panel, $\omega_2=1$ is fixed while $\omega_1\rightarrow\pm\infty$; in the right panel, $\omega_1=1$ is fixed while $\omega_2\rightarrow\pm\infty$. The black dashed lines indicate the directional exponential decay rates predicted by the full-ETH bound~\eqref{eq:QFT-general-coordinate-bound}. In both panels, the numerical asymptotics are consistent with these predicted rates.}
    \label{fig:kappa3}
\end{figure}

\paragraph{Third free cumulant.}
The multitime Fourier transform of the third-order cumulant is defined as
\begin{equation} \label{eq:Fourierk3}
    \hat{\kappa}_3(\omega_1,\omega_2)=\int_{-\infty}^{\infty}\frac{dt_1}{2\pi} \int_{-\infty}^{\infty}\frac{dt_2}{2\pi} \, e^{- i\omega_1 t_1-i \omega_2 t_2} \kappa_3^{\textrm{reg}}(t_1,t_2)
\end{equation}
Writing the Fourier representation of each individual kernel appearing in $\kappa_3^{\textrm{reg}}(t_1,t_2)$ as
\begin{align}
  &K_{\delta_{12},\beta/3}(t_1-t_2)= \int_{-\infty}^{\infty}d\omega\, e^{i \omega (t_1-t_2)} K_{\delta_{12},\beta/3}(\omega)\,, \nonumber \\
 & K_{\delta_{23},\beta/3}(t_2)= \int_{-\infty}^{\infty}d\omega'\, e^{i \omega' t_2} K_{\delta_{23},\beta/3}(\omega')\,,\nonumber \\
&  K_{\delta_{31},2\beta/3}(t_1)= \int_{-\infty}^{\infty}d\omega''\, e^{i \omega'' t_1} K_{\delta_{31},2\beta/3}(\omega'')\,.
\end{align}
and substituting these expressions into Eq.~\eqref{eq:Fourierk3}, we obtain
\begin{equation}
\hat{\kappa}_3(\omega_1,\omega_2)=\frac{c_{123}}{2\pi}  \left(\frac{\pi}{\beta}\right)^{\Delta_1+\Delta_2+\Delta_3} \int_{-\infty}^{\infty}d\omega \, \hat{K}_{\delta_{12},\beta/3}(\omega)\, \hat{K}_{\delta_{23},\beta/3}(\omega_2+\omega)\,\hat{K}_{\delta_{31},2\beta/3}(\omega_1-\omega)
\end{equation}
This convolution formula, together with the Fourier transform of the shifted kernel in Eq.~\eqref{eq:FourierKernel}, allows us to study numerically the large-frequency behavior of $\hat{\kappa}_3(\omega_1,\omega_2)$ and compare it with the directional decay rates predicted by the full-ETH bounds. Figure~\ref{fig:kappa3} illustrates this behavior with $\omega_2$ fixed and $|\omega_1|$ large in the left panel, and with $\omega_1$ fixed and $|\omega_2|$ large in the right panel. In both cases, the asymptotic behavior of the Fourier transform realizes the limiting exponential decay rates predicted by Eq.~\eqref{eq:QFT-general-coordinate-bound}.

\subsection{Fourth-order free cumulants in holographic CFTs}
\label{sec:kappa4}

In this subsection, we study the multifrequency behavior of holographic
fourth-order free cumulants. We first explain why, for the operator
configuration considered below, the fourth free cumulant is precisely
the combination naturally computed by the holographic eikonal
prescription.

Consider two centered scalar primaries $V$ and $W$, with dimensions
$\Delta_V$ and $\Delta_W$. We place the two $V$ insertions at
$\mathbf{x}_1$ and the two $W$ insertions at $\mathbf{x}_2$, and
introduce the shorthand
\begin{equation}
V_i
\equiv
V(z_i,\mathbf{x}_1)\,,
\qquad
i=1,3\,,
\qquad
W_j
\equiv
W(z_j,\mathbf{x}_2)\,,
\qquad
j=2,4\,.
\end{equation}
For arbitrary complex insertion times $\vec z=(z_1,z_2,z_3,z_4)$, we
consider the thermal four-point function
\begin{equation}
F_4(\vec z;\mathbf{x}_1,\mathbf{x}_2)
=
\left\langle
V_1\,W_2\,V_3\,W_4
\right\rangle_\beta\,.
\label{eq:holo-four-point}
\end{equation}
The ordering of the operators is fixed as written, while their
imaginary parts specify the corresponding thermal contour. For centered operators, the decomposition of this correlator in terms
of free cumulants gives
\begin{equation}
\kappa_4(1,2,3,4)
=
F_4(1,2,3,4)
-
F_{12}\,F_{34}
-
F_{14}\,F_{23}\,,
\label{eq:kappa4-otoc}
\end{equation}
where $F_{ij}$ denotes the thermal two-point function between the
operators at positions $i$ and $j$, evaluated with the ordering and
imaginary-time separation inherited from the original four-point
function.

We take $V$ and $W$ to be orthogonal primaries such that all mixed two-point
functions vanish:
\begin{equation}
F_{12}
=
F_{34}
=
F_{14}
=
F_{23}
=
0\,.
\label{eq:mixed-two-point-functions}
\end{equation}
It follows immediately that
\begin{equation}
\kappa_4(1,2,3,4)
=F_4(1,2,3,4)
\,.
\label{eq:kappa4-equals-F4}
\end{equation}
The ordinary connected correlator is instead obtained by subtracting
all three pairwise factorizations:
\begin{equation}
F_{\mathrm{conn}}(1,2,3,4)
=
F_4(1,2,3,4)
-
F_{12}\,F_{34}
-
F_{14}\,F_{23}
-
F_{13}\,F_{24}\,.
\end{equation}
Using Eq.~\eqref{eq:mixed-two-point-functions}, this reduces to
\begin{equation}
F_{\mathrm{conn}}(1,2,3,4)
=
F_4(1,2,3,4)
-
F_{13}\,F_{24}\,,
\end{equation}
where
\begin{equation}
F_{13}
=
\left\langle V_1V_3\right\rangle_\beta\,,
\qquad
F_{24}
=
\left\langle W_2W_4\right\rangle_\beta\,.
\end{equation}
Consequently,
\begin{equation}
\kappa_4(1,2,3,4)
=
F_{\mathrm{conn}}(1,2,3,4)
+
F_{13}\,F_{24}
=
F_4(1,2,3,4)
\,.
\label{eq:kappa4-connected-crossing}
\end{equation}

The holographic eikonal prescription naturally computes precisely this
combination. In the absence of gravitational scattering, $g=0$, it
reduces to the crossing factorization
\begin{equation}
F_{13}\,F_{24}
=
\left\langle V_1V_3\right\rangle_\beta\,
\left\langle W_2W_4\right\rangle_\beta\,,
\end{equation}
whereas the nontrivial eikonal phase generates the ordinary connected
gravitational contribution. Having established this identification, we
now specialize the general holographic result
\eqref{eq:higher-dimensional-resolved-otoc} to the fourth-order free
cumulant with equally spaced insertions on the thermal circle. Writing
the complex insertion times as
\begin{equation}
z_i
=
\frac{2\pi}{\beta}\,t_i
+
i\,\epsilon_i\,,
\end{equation}
the equally spaced configuration corresponds to
\begin{equation}
\epsilon_1=0\,,
\qquad
\epsilon_2=\frac{\pi}{2}\,,
\qquad
\epsilon_3=\pi\,,
\qquad
\epsilon_4=\frac{3\pi}{2}\,.
\label{eq:equally-spaced-regulators}
\end{equation}
Substituting this configuration into the general holographic
prescription applied for Rindler-AdS$_{d+1}$/CFT$_{d}$, we obtain\footnote{We absorb $a_0$ into the definition of $g$.}
\begin{align}
\kappa_4(\vec t;\mathbf{x}_1,\mathbf{x}_2)
={}&
\mathcal N\,
\mathcal P_{13,\mathrm{eq}}^{(V)}(\vec t)\,
\mathcal P_{24,\mathrm{eq}}^{(W)}(\vec t)
\int_{H^{d-1}}
d\mathbf{x}\,d\mathbf{y}
\int_0^\infty
dp\,dq\,
p^{2\Delta_V-1}\,
q^{2\Delta_W-1}
\nonumber\\
&\times
e^{-p\cosh d(\mathbf{x},\mathbf{x}_1)}
e^{-q\cosh d(\mathbf{y},\mathbf{x}_2)}
\exp\left[
-\frac{
g\,p\,q\,h(\mathbf{x},\mathbf{y})
}{
\mathcal D(\vec t)
}
\right]\,.
\label{eq:equally-spaced-holographic-kappa4}
\end{align}
where
\begin{align}
\mathcal P_{13,\mathrm{eq}}^{(V)}(\vec t)
&=
\left[
2i\,
\cosh\left(
\frac{\pi}{\beta}(t_3-t_1)
\right)
\right]^{-2\Delta_V}\,,
\\
\mathcal P_{24,\mathrm{eq}}^{(W)}(\vec t)
&=
\left[
2i\,
\cosh\left(
\frac{\pi}{\beta}(t_4-t_2)
\right)
\right]^{-2\Delta_W}\,,\\
\mathcal D(\vec t)
&=
\left(
e^{\frac{2\pi}{\beta}t_3}
+
e^{\frac{2\pi}{\beta}t_1}
\right)
\left(
e^{-\frac{2\pi}{\beta}t_4}
+
e^{-\frac{2\pi}{\beta}t_2}
\right)\,.
\end{align}
Note that the eikonal exponent becomes purely real for equally spaced insertions on the thermal circle. 

It is convenient to introduce new time variables adapted to the two operator pairs and to their relative boost. We define
\begin{equation}
t_u=\frac{t_3-t_1}{2}\,,
\qquad
t_v=\frac{t_2-t_4}{2}\,,
\qquad
t_s=\frac{t_2+t_4-t_1-t_3}{2}\,.
\label{eq:tu-tv-ts}
\end{equation}
The variables $t_u$ and $t_v$ measure the relative time separations within the $V$ and $W$ pairs, respectively, while $t_s$ measures the relative separation between the centers of the two pairs.
In the Regge limit, $t_s$ therefore plays the role of the relative boost time between the two scattering states.

The dependence on the transverse coordinates can be conveniently
collected into a transverse eikonal kernel. We define
\begin{equation}
\mathcal{K}_d(p,q;\lambda;\mathbf{x}_1,\mathbf{x}_2)
\equiv
\int_{H^{d-1}} d\mathbf{x}\,d\mathbf{y}\,
\exp\left[
-p\cosh d(\mathbf{x},\mathbf{x}_1)
-q\cosh d(\mathbf{y},\mathbf{x}_2)
-\lambda\,p q\,h(\mathbf{x},\mathbf{y})
\right]\,,
\label{eq:transverse-kernel}
\end{equation}
In terms of the variables introduced in
Eq.~\eqref{eq:tu-tv-ts}, the time-dependent effective eikonal
coupling is
\begin{equation}
\lambda(t_u,t_v,t_s)
=
\frac{g\,e^{\kappa t_s}}
{4\cosh(\kappa t_u)\cosh(\kappa t_v)}\,,
\qquad
\kappa=\frac{2\pi}{\beta}\,.
\label{eq:effective-eikonal-coupling}
\end{equation}
Thus, all dependence on the transverse geometry is encoded in
$\mathcal{K}_d$, while the relative Regge boost enters through the
factor $e^{\kappa t_s}$ in $\lambda$. In terms of these new variables the thermal free cumulant becomes
\begin{align}
\kappa_4(t_u,t_v,t_s)
={}&
\mathcal N\,
\left[2i\cosh(\kappa t_u)\right]^{-2\Delta_V}
\left[2i\cosh(\kappa t_v)\right]^{-2\Delta_W}
\nonumber\\
&\times
\int_0^\infty dp\,dq\,
p^{2\Delta_V-1}q^{2\Delta_W-1}
\mathcal K_d
\left(
p,q;\lambda(t_u,t_v,t_s);
\mathbf{x}_1,\mathbf{x}_2
\right)\,.
\end{align}

We now consider the multitime Fourier transform of the fourth
free cumulant. Using time-translation invariance, we set $t_4=0$ and
define
\begin{equation}
\widehat{\kappa}_4(\omega_1,\omega_2,\omega_3)
=
\int_{-\infty}^{\infty}dt_1\,dt_2\,dt_3\,
e^{-i(\omega_1t_1+\omega_2t_2+\omega_3t_3)}
\kappa_4(t_1,t_2,t_3,0)\,,
\label{eq:fourier-kappa4-higher-d}
\end{equation}
where the fourth frequency is fixed by energy conservation $\omega_4=-\omega_1-\omega_2-\omega_3$. 
In terms of the variables defined in
Eq.~\eqref{eq:tu-tv-ts}, the multitime Fourier transform then takes the simple form
\begin{equation}
\widehat{\kappa}_4(\vec\omega)
=
4
\int_{-\infty}^{\infty}
dt_u\,dt_v\,dt_s\,
e^{i\Omega_u t_u-i\Omega_v t_v+i\Omega_s t_s}
\kappa_4(t_u,t_v,t_s)\,.
\label{eq:fourier-kappa4-tuvs}
\end{equation}
where
\begin{equation}
\Omega_u=\omega_1-\omega_3\,,
\qquad
\Omega_v=\omega_2-\omega_4\,,
\qquad
\Omega_s=\omega_1+\omega_3
=-(\omega_2+\omega_4)\,.
\label{eq:conjugate-frequencies-tuvs}
\end{equation}
In particular, $\Omega_s=\omega_1+\omega_3$ is conjugate to the
relative boost time $t_s$. This will be the natural frequency variable
controlling the Regge contribution to the Fourier transform.

Using the variables introduced above, the multitime Fourier transform can be written as
\begin{align}
\widehat{\kappa}_4(\vec{\omega})
={}&
4\mathcal N
\int_{H^{d-1}}d\mathbf{x}\,d\mathbf{y}
\int_0^\infty dp\,dq\,
p^{2\Delta_V-1}q^{2\Delta_W-1}
e^{-p\cosh d(\mathbf{x},\mathbf{x}_1)}
e^{-q\cosh d(\mathbf{y},\mathbf{x}_2)}
\nonumber\\
&\times
\int_{-\infty}^{\infty}dt_u\,
\int_{-\infty}^{\infty}dt_v\,
\int_{-\infty}^{\infty}dt_s\,
e^{i\Omega_u t_u-i\Omega_v t_v+i\Omega_s t_s}
\nonumber\\
&\times
\left[2i\cosh(\kappa t_u)\right]^{-2\Delta_V}
\left[2i\cosh(\kappa t_v)\right]^{-2\Delta_W}
\exp\left[
-\frac{g\,pq\,h(\mathbf{x},\mathbf{y})}
{4\cosh(\kappa t_u)\cosh(\kappa t_v)}
e^{\kappa t_s}
\right]\,,
\label{eq:fourier-kappa4-tuvs-full}
\end{align}
The $t_s$ dependence is entirely contained in the relative Regge-boost
factor. At fixed $t_u,t_v,p,q,\mathbf{x},\mathbf{y}$, the corresponding
integral is
\begin{equation}
I_s
=
\int_{-\infty}^{\infty}dt_s\,
e^{i\Omega_s t_s}
e^{-H e^{\kappa t_s}}\,,
\qquad
H=
\frac{g\,pq\,h(\mathbf{x},\mathbf{y})}
{4\cosh(\kappa t_u)\cosh(\kappa t_v)}\,.
\label{eq:ts-integral}
\end{equation}
Introducing
\begin{equation} \label{eq:nu}
\nu
=
\frac{\Omega_s}{\kappa}
=
\frac{\omega_1+\omega_3}{\kappa}\,,
\end{equation}
and making the change of variables $z=H e^{\kappa t_s}$, one obtains\footnote{We understand this Fourier transform distributionally, by inserting the exponential $e^{\epsilon\kappa t_s}$ and taking $\epsilon\to0^+$. For $\nu\neq0$, this gives the current expression. The zero frequency contribution does not affect the large-$|\nu|$ asymptotics considered below.}
\begin{equation}
    I_s=\frac{H^{-i\nu}}{\kappa}\Gamma(i \nu)\,.
\end{equation}
Plugging this result into the expression for the cumulant, we obtain
\begin{align}
\widehat{\kappa}_4(\vec{\omega})
={}&
\frac{4\mathcal N}{\kappa}
\left(\frac{g}{4}\right)^{-i\nu}
\Gamma(i\nu)
\int_{H^{d-1}}d\mathbf{x}\,d\mathbf{y}\,
h(\mathbf{x},\mathbf{y})^{-i\nu}
\nonumber\\
&\times
\int_0^\infty dp\,dq\,
p^{2\Delta_V-1-i\nu}
q^{2\Delta_W-1-i\nu}
e^{-p\cosh d(\mathbf{x},\mathbf{x}_1)}
e^{-q\cosh d(\mathbf{y},\mathbf{x}_2)}
\nonumber\\
&\times
\int_{-\infty}^{\infty}dt_u\,
e^{i\Omega_u t_u}
\left[2i\cosh(\kappa t_u)\right]^{-2\Delta_V}
\left[\cosh(\kappa t_u)\right]^{i\nu}
\nonumber\\
&\times
\int_{-\infty}^{\infty}dt_v\,
e^{-i\Omega_v t_v}
\left[2i\cosh(\kappa t_v)\right]^{-2\Delta_W}
\left[\cosh(\kappa t_v)\right]^{i\nu}\,,
\label{eq:kappa4-after-ts}
\end{align}
The integral in $p$ gives
\begin{equation}
\int_0^\infty dp\,
p^{2\Delta_V-1-i\nu}
e^{-p\cosh d(\mathbf{x},\mathbf{x}_1)}
=
\Gamma(2\Delta_V-i\nu)
\left[\cosh d(\mathbf{x},\mathbf{x}_1)\right]^{-2\Delta_V+i\nu}\,,
\end{equation}
with an analogous result for the integral in $q$. Substituting the above results in the expression for $\widehat{\kappa}_4(\vec{\omega})$, we find
\begin{align}
\widehat{\kappa}_4(\vec{\omega})
={}&
\frac{4\mathcal N}{\kappa}
\left(\frac{g}{4}\right)^{-i\nu}
\Gamma(i\nu)
\Gamma(2\Delta_V-i\nu)
\Gamma(2\Delta_W-i\nu)
\nonumber\\
&\times
\int_{H^{d-1}}d\mathbf{x}\,d\mathbf{y}\,
h(\mathbf{x},\mathbf{y})^{-i\nu}
\left[\cosh d(\mathbf{x},\mathbf{x}_1)\right]^{-2\Delta_V+i\nu}
\left[\cosh d(\mathbf{y},\mathbf{x}_2)\right]^{-2\Delta_W+i\nu}
\nonumber\\
&\times
(2i)^{-2\Delta_V}
\int_{-\infty}^{\infty}dt_u\,
e^{i\Omega_u t_u}
\left[\cosh(\kappa t_u)\right]^{-2\Delta_V+i\nu}
\nonumber\\
&\times
(2i)^{-2\Delta_W}
\int_{-\infty}^{\infty}dt_v\,
e^{-i\Omega_v t_v}
\left[\cosh(\kappa t_v)\right]^{-2\Delta_W+i\nu}\,.
\label{eq:kappa4-after-pq}
\end{align}
The remaining integrals over $t_u$ and $t_v$ can be evaluated using the relation
\begin{equation}
\int_{-\infty}^{\infty}dt\,
e^{i\Omega t}
[\cosh(\kappa t)]^{-a}
=
\frac{2^{a-1}}{\kappa}
\frac{
\Gamma\left(\frac{a}{2}+\frac{i\Omega}{2\kappa}\right)
\Gamma\left(\frac{a}{2}-\frac{i\Omega}{2\kappa}\right)
}{
\Gamma(a)
}\,.
\end{equation}
Using the above relation for the integrals over $t_u$ and $t_v$, and
expressing $\Omega_u$ and $\Omega_v$ in terms of $\omega_1,\omega_2$,
and $\omega_3$ using Eq.~\eqref{eq:conjugate-frequencies-tuvs}, together
with Eq.~\eqref{eq:nu}, we find\footnote{A similar frequency-space expression for
observer-dressed OTOCs in the de Sitter static patch was recently
obtained in Ref.~\cite{Chen:2026boh}.}
\begin{align}
\widehat{\kappa}_4(\vec{\omega})
={}&
\frac{\mathcal N}{\kappa^3}\,
i^{-2(\Delta_V+\Delta_W)}
g^{-i\nu}\,
\Gamma(i\nu)
\Gamma\left(\Delta_V-\frac{i\omega_1}{\kappa}\right)
\Gamma\left(\Delta_V-\frac{i\omega_3}{\kappa}\right)
\nonumber\\
&\times
\Gamma\left(\Delta_W+\frac{i\omega_2}{\kappa}\right)
\Gamma\left(\Delta_W+\frac{i\omega_4}{\kappa}\right)\mathcal{B}_d(\nu) 
\label{eq:fourier-kappa4-higher-dimensional}
\end{align}
where $\omega_4=-\omega_1-\omega_2-\omega_3$ and
\begin{equation}
    \mathcal{B}_d(\nu)=\int_{H^{d-1}}d\mathbf{x}\,d\mathbf{y}\,
h(\mathbf{x},\mathbf{y})^{-i\nu}
\left[\cosh d(\mathbf{x},\mathbf{x}_1)\right]^{-2\Delta_V+i\nu}\,\left[\cosh d(\mathbf{y},\mathbf{x}_2)\right]^{-2\Delta_W+i\nu}\,.
\end{equation}

\paragraph {Lower dimensional case.} The lower-dimensional AdS$_2$ result can be obtained from the expression above by simply setting $\mathcal{B}_b(\nu)=1$. We first analyze this lower-dimensional case and then discuss the subtleties that arise in higher dimensions.

To study the large-frequency behavior of the expression, it is useful to employ the Stirling asymptotic formula for the Gamma function,
\begin{equation}
|\Gamma(a+i\nu)|
\sim
\sqrt{2\pi}\,
|\nu|^{a-\frac12}
e^{-\frac{\pi}{2}|\nu|},
\qquad
|\nu|\rightarrow\infty.
\end{equation}
Applying this formula to the frequency-dependent Gamma functions in the expression above, while keeping the remaining independent frequencies fixed, we find
\begin{equation}
\left|\widehat{\kappa}_4(\vec{\omega})\right|
\sim
\operatorname{poly}(W)
\begin{cases}
e^{-3\beta W/4}\,, & \omega_1=\pm W\,,\\[2pt]
e^{-\beta W/2}\,, & \omega_2=\pm W\,,\\[2pt]
e^{-3\beta W/4}\,, & \omega_3=\pm W\,,
\end{cases}
\qquad W\rightarrow\infty\,.
\label{eq:kappa4-coordinate-asymptotics}
\end{equation}
Comparing with Eq.~\eqref{eq:QFT-general-coordinate-bound}, the
strongest coordinate bounds are therefore saturated for
$\omega_1\rightarrow-\infty$, $\omega_2\rightarrow\pm\infty$, and
$\omega_3\rightarrow+\infty$, while the opposite $\omega_1$ and
$\omega_3$ directions decay faster than required. This behavior follows from the symmetry of the holographic result,
\begin{equation}
    \widehat{\kappa}_4(-\vec{\omega})
    =
    \widehat{\kappa}_4(\vec{\omega})^*
    \qquad \Longrightarrow \qquad
    \left|\widehat{\kappa}_4(-\vec{\omega})\right|
    =
    \left|\widehat{\kappa}_4(\vec{\omega})\right|\,.
\end{equation}
Consequently, once the stronger bound is saturated in one frequency
direction, the opposite direction must exhibit the same exponential
decay rate, and therefore decays faster than required by the weaker
bound.

\paragraph{Higher-dimensional case.}
The analysis in higher dimensions is more subtle because the transverse
dynamics introduces the additional frequency-dependent factor
$\mathcal{B}_d(\nu)$, with $\nu=(\omega_1+\omega_3)/\kappa$. Consider first $\omega_2\rightarrow\pm\infty$, with $\omega_1$ and
$\omega_3$ fixed. In this limit $\nu$ remains fixed, and therefore
$\mathcal{B}_d(\nu)$ contributes only an $\omega_2$-independent factor.
Consequently, the bounds in the $\omega_2$ directions remain saturated.
For $\omega_1$ or $\omega_3$ large, by contrast, one has
$|\nu|\sim |\omega_{1,3}|/\kappa$, and the asymptotic behavior of
$\mathcal{B}_b(\nu)$ can modify the decay inherited from the Gamma
functions, and in principle may lead to a non-saturation of the frequency bounds predicted by full ETH.

\begin{figure}
    \centering
    \includegraphics[width=0.6\linewidth]{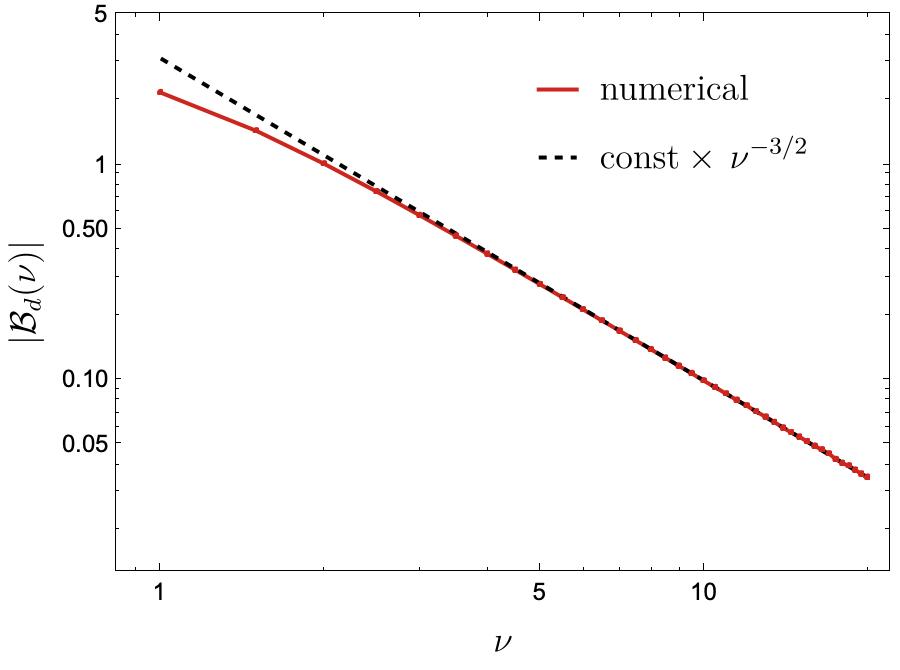}
    \caption{Large-$|\nu|$ behavior of the transverse factor
    $|\mathcal{B}_b(\nu)|$ in Rindler--AdS$_3$/CFT$_2$ for
    $\Delta_V=\Delta_W=1$ and $b=|x_1-x_2|=1$. The numerical result is compared
    with the asymptotic power law $\text{const} \times|\nu|^{-3/2}$ (black dashed line).}
    \label{fig:Bb-AdS3}
\end{figure}

To investigate this possibility, we numerically study the large frequency behavior of $\mathcal{B}_b(\nu)$ for Rindler-AdS$_{3}$, in which case one has $h(x,y)\propto e^{-|x-y|}$.
For $\Delta_V=\Delta_W=1$ and $b=|x_1-x_2|=1$, the numerical evaluation shown in
Fig.~\ref{fig:Bb-AdS3} exhibits the power-law behavior
\begin{equation}
|\mathcal{B}_b(\nu)|
\sim
|\nu|^{-3/2}\,,
\qquad
|\nu|\gg1\,.
\end{equation}
Thus, in Rindler--AdS$_3$ the transverse contribution changes only the
power-law prefactor of the Fourier transform and does not introduce any
additional exponential suppression. The exponential large-frequency decay
therefore remains entirely controlled by the Gamma functions, and the
strongest full-ETH bounds remain saturated. The investigation of the behavior of $|\mathcal{B}_b(\nu)|$ for $d>2$ (AdS$_4$ or higher) is numerically more demanding and we leave it for future work.

\subsection{Subtlety about time ordering}

There is a subtlety in applying the eikonal expression inside a multitime Fourier transform. Throughout the transform, the imaginary positions of the operators along the thermal circle are held fixed at equally spaced separations, so that the prescribed OTO contour is preserved, while their real insertion times are integrated over the entire real line. The eikonal correlator is derived in the gravitational Regge regime, corresponding to a parametrically large relative boost between the two pairs of operators, whereas the Fourier transform also samples configurations outside this regime. Therefore, when integrating over all times, we are extrapolating the eikonal formula beyond its strict domain of validity. For the large-frequency asymptotics of interest here, however, the dominant region of the Fourier integral corresponds to increasingly large relative boosts. Thus, the large-frequency behavior is naturally associated with the Regge limit, providing an a posteriori justification for the use of the eikonal approximation.

This can already be seen from the integral over $t_s$, which gives
\begin{equation}
\label{eq:ts-integral2}
I(\nu)
=
\int_{-\infty}^{\infty}dt_s\,
e^{i\kappa\nu t_s}
\exp\left[-H e^{\kappa t_s}\right]
=
\frac{H^{-i\nu}}{\kappa}\Gamma(i\nu)\,.
\end{equation}
The asymptotic expression for the Gamma function produces the factor
$\operatorname{poly}(|\nu|)e^{-\pi|\nu|/2}$, which ultimately leads to
the saturation of the full-ETH bounds when combined with the other
Gamma functions. We note that the same asymptotic behavior can also be
understood directly from a steepest-descent approximation of the
$t_s$ integral in Eq.~\eqref{eq:ts-integral2}. Defining the exponent
\begin{equation}
\Phi(t_s)
=
i\kappa\nu t_s-H e^{\kappa t_s}\,,
\end{equation}
the saddle condition $\Phi'(t_{s,*})=0$ gives
\begin{equation}
H e^{\kappa t_{s,*}}
=
i\nu\,.
\end{equation}
For real nonzero $\nu$, the relevant saddle can be written as
\begin{equation}
t_{s,*}
=
\frac{1}{\kappa}\log\frac{|\nu|}{H}
+
i\,\frac{\pi}{2\kappa}\,\operatorname{sgn}(\nu)\,.
\end{equation}
Therefore, the large-$|\nu|$ implies that the real part of $t_{s,*}$ is also very large, so that the $t_s$ integral is controlled by the Regge regime in which the eikonal approximation is well justified.

\section{Conclusions and Future Directions} \label{sec-conclusions}
In this work, we have studied full ETH in the context of quantum field theories (QFTs). Full ETH introduces a hierarchy of smooth ETH functions that encode correlations among multiple products of operator matrix elements in the energy eigenbasis. When full ETH holds, these functions are directly related to the multitime Fourier transforms of thermal free cumulants. Previous works derived large-frequency bounds on these functions from the finiteness of free cumulants at equal times \cite{Murthy:2019fgs,Pappalardi:2022aaz}. Such derivations are naturally suited to lattice systems, since in continuum QFT equal-time correlators generally suffer from ultraviolet singularities.

To extend these bounds to QFTs, we instead consider regulated correlators with operator insertions equally spaced along the thermal circle. We show that their analyticity in complex time, together with a boundedness assumption on the time integral of the shifted free cumulants and at-most power-law growth near the short-distance boundaries of the analyticity domain, leads to analogous large-frequency bounds in continuum QFT. These bounds exhibit an asymmetric dependence on the different frequency directions, which can be understood geometrically from the allowed displacements of the operator insertions along the thermal circle after fixing the last operator at $t_n=0$. The same asymmetric structure is already present for lattice systems, although, to our knowledge, this directional dependence has not been emphasized in previous discussions of full-ETH bounds. We note that the analyticity and boundedness assumptions entering this argument are similar in spirit to those appearing in the derivation of the conventional chaos bound \cite{Maldacena:2015waa}.

A practical way to test these bounds in QFTs is to compute directly the multitime Fourier transforms of thermal free cumulants. If full ETH holds, the resulting large-frequency behavior should be consistent with the bounds derived above. Such a test does not by itself establish full ETH, but it provides a nontrivial consistency check on the expected high-frequency structure of the corresponding ETH smooth functions.
We tested these bounds in thermal CFTs. At second order, the thermal two-point function saturates the bound, as expected from the standard KMS analytic structure. The three-point function also saturates the corresponding bounds and provides a simple realization of their asymmetric dependence on the different frequencies entering the multitime Fourier transform. Finally, we studied the fourth-order free cumulant in holographic CFTs. In this case, the leading gravitational Regge contribution realizes the strongest full-ETH decay rate associated with each independent frequency. Owing to the symmetry of the gravitational free cumulant, the same decay is obtained in the opposite frequency direction whenever the general full-ETH bound is weaker. In particular, the $\omega_1$ and $\omega_3$ tails decay as $e^{-3\beta W/4}$ in both directions, while the $\omega_2$ tails decay as $e^{-\beta W/2}$. Thus, the gravitational result saturates the strongest bounds associated with all independent frequencies. This provides a genuinely dynamical test of the bounds beyond the kinematically fixed two- and three-point functions.

The full-ETH frequency bound for the fourth-order free cumulant is closely related to the conventional chaos bound. As pointed out in Ref.~\cite{Murthy:2019fgs}, the large-frequency constraint can be converted into a bound on the Lyapunov exponent. In particular, starting from the holographic form of the OTOC and replacing the maximal Lyapunov exponent by a general growth rate $\lambda$, requiring its Fourier transform to satisfy the full-ETH frequency bound leads to $\lambda\leq 2\pi/\beta$, where $\beta$ is the inverse temperature of the thermal state. In this sense, the full-ETH bounds may be viewed as frequency-space counterparts of the chaos bound, with the additional feature that they extend beyond the fourth-order cumulant to a complete hierarchy of higher-order free cumulants. It is then natural to ask whether holographic systems also saturate the corresponding higher-order bounds. A possible first step would be to study the free cumulants associated with six-point OTOCs, such as those considered in Ref.~\cite{Haehl:2021tft}.

This higher-order perspective is also naturally connected to the gravitational scrambling algebra introduced in Ref.~\cite{Penington:2025hrc}. In this framework, sufficiently late-time operators become free from early-time operators, and freeness itself is characterized by the vanishing of mixed free cumulants at all orders. It would therefore be interesting to determine whether this algebraic structure is consistent with the full-ETH bounds not only through fourth order, as tested here, but at higher orders as well. More generally, the bounds \eqref{eq:general-coordinate-bound} provide a systematic framework for investigating higher-order aspects of scrambling in both lattice models and QFTs.

We conclude by pointing out a few possible directions for future work. Throughout this work, we have considered correlators with operator insertions equally spaced along the thermal circle. It would be interesting to derive the corresponding frequency bounds for other choices of thermal regularization, in the spirit of Ref.~\cite{Romero-Bermudez:2019vej}, and to compare the resulting large-frequency behavior with the regularization dependence of the corresponding Lyapunov exponents. Another natural direction is to study the frequency-space behavior of free cumulants in models that do not exhibit maximal chaos and investigate how their asymptotic decay is related to the corresponding Lyapunov exponent. A promising setting for this analysis is the effective theory of scramblons proposed in Ref.~\cite{Gu:2021xaj}. We expect to report on this in future work.

\section*{Acknowledgments}
We thank Bartek Czech, Jan de Boer, Ben Freivogel, Mark Mezei, Miguel Tierz, and Huajia Wang for insightful discussions. R.~Esp\'indola thanks Roberto Emparan for hospitality and useful discussions at the Institute of Cosmos Sciences of the University of Barcelona (ICCUB), where part of this work was carried out. He also thanks the organizers of “Holo-Asia 2026'' (Jeju) and of the “Amsterdam String Summer Workshop 2026''. R.~Esp\'indola is supported by the Shuimu Tsinghua Scholar Program. V.~Jahnke was supported by the Conselho Nacional de Desenvolvimento Científico e Tecnológico (CNPq), Brazil, under grant Processo 446326/2024-0 (Bolsa Conhecimento Brasil – BCB-1).
%%%%%%%%%%%%%%%%%%%%%%%%%%%%%%%%%%%%%%%%%%%%%%%%%%
\appendix
\se{Classical Probability Theory} \label{app:ClassicalProbability}
In this section, we provide a brief review of classical probability theory. Our main goal is to introduce classical cumulants and a diagrammatic method based on partitions that clarifies the combinatorial structure underlying the relation between moments and cumulants.
For a more in-depth treatment of the formal aspects of probability theory, we refer to \cite{Tao2010ProbNotes}, while a detailed account of the combinatorial techniques underlying the computation of classical moments can be found, for example, in \cite{novak2012lecturesfreeprobability,SpeicherNica2006}.

\paragraph{Definition 1: Probability space.}
A probability space is a triple, also known as a Kolmogorov triple,
$(\Omega, \Sigma, \mu)$, consisting of a sample space $\Omega$ (a set),
a $\sigma$-algebra $\Sigma$ of measurable subsets of $\Omega$ (events), and a
probability measure $\mu : \Sigma \to [0,1]$ satisfying:
\begin{itemize}
    \item[(i)] \textbf{Non-negativity:} $\mu(A) \geq 0$ for all $A \in \Sigma$;
    \item[(ii)] \textbf{Normalization:} $\mu(\Omega) = 1$;
    \item[(iii)] \textbf{$\sigma$-additivity:} for any countable collection of pairwise disjoint sets $\{A_i\}_{i=1}^\infty \subset \Sigma$,
    \begin{equation}
        \mu\!\left(\bigcup_{i=1}^{\infty} A_i\right)
        = \sum_{i=1}^{\infty} \mu(A_i).
    \end{equation}
\end{itemize}

A simple example of a probability space is the experiment of tossing a fair coin, which may result in heads (H) or tails (T). In this case, the sample space is $\Omega = \{H, T\}$, and the $\sigma$-algebra of all measurable subsets is
$
\Sigma = \{\varnothing, \{H\}, \{T\}, \Omega\}.
$
The probability measure $\mu$ assigns probabilities to each event as
\begin{equation}
   \mu(\varnothing) = 0, \quad
\mu(\{H\}) = \tfrac{1}{2}, \quad
\mu(\{T\}) = \tfrac{1}{2}, \quad
\mu(\Omega) = 1. 
\end{equation}

\paragraph{Definition 2: Independence.}
Given a probability space $(\Omega,\Sigma,\mu)$, two events
$A,B \in \Sigma$ are said to be independent if
\begin{equation}
    \mu(A \cap B) = \mu(A)\mu(B).
\end{equation}
As an example, consider the experiment of tossing a fair coin twice. In this case, the sample space is $\Omega = \{HH, HT, TH, TT\},$ and $\Sigma$ consists of all subsets of $\Omega$, with each outcome having probability $1/4$. Let $A = \{\text{first toss is } H\} = \{HH, HT\}$ and
$B = \{\text{second toss is } H\} = \{HH, TH\}$. Then,
\begin{equation}
    \mu(A) = \frac{1}{2}, \quad \mu(B) = \frac{1}{2}, \quad \mu(A \cap B) = \mu(\{HH\}) = \frac{1}{4},
\end{equation}
so that $\mu(A \cap B) = \mu(A)\mu(B)$, and hence $A$ and $B$ are independent, reflecting the fact that the outcome of the first toss does not influence the outcome of the second.

\paragraph{Definition 3: Random variable.}
A real-valued random variable is a measurable function
$X : \Omega \to \mathbb{R}$, i.e.\ such that for every Borel measurable set
$B \subset \mathbb{R}$, the preimage $X^{-1}(B) \in \Sigma$ is an event. This condition ensures that statements of the form $X \in B$ correspond to measurable events, with probabilities given by $\mathrm{Prob}(X \in B) := \mu(X^{-1}(B))$, which defines the pushforward measure $\mu_X(B) = \mathrm{Prob}(X \in B)$.

\paragraph{Radon--Nikodym theorem: probability density function.}
Given a real-valued random variable $X$ and a reference measure $\nu$ on $\mathbb{R}$ (for instance, the Lebesgue measure), if the distribution $\mu_X$ of $X$ is absolutely continuous with respect to $\nu$, then there exists a non-negative function $f \in L^1(\mathbb{R}, d\nu)$ such that
\begin{equation}
   \text{Prob}(X \in A)= \mu_X(A) = \int_A f(x)\, d\nu(x),
\end{equation}
for all measurable sets $A \subset \mathbb{R}$. 
The function $f$ is called the \emph{probability density function} (PDF) of $X$, and is normalized so that $\int_{\mathbb{R}} f(x)\, d\nu(x) = 1$. In the case where $\nu$ is the Lebesgue measure, this reduces to the familiar expression $\mu_X(A) = \int_A f(x)\, dx$.

For certain classes of probability density functions, knowing the distribution of $X$ is equivalent to knowing all the order-$n$ moments $m_n(X)$, also denoted as $\mathbb{E}[X^n]$, which are defined as follows:
\begin{equation}
    m_n(X) = \mathbb{E}[X^n] = \int_{\mathbb{R}} x^n f(x)\, dx\,.
\end{equation}
It is convenient to introduce the moment-generating function $M_X(t)$, defined by
\begin{equation}
    M_X(t) = \mathbb{E}[e^{tX}] = \int_{\mathbb{R}} e^{t x} f(x)\, dx\,,
\end{equation}
which encodes all moments of the distribution through its Taylor expansion around $t=0$,
\begin{equation}
    M_X(t) = \sum_{n=0}^{\infty} \frac{m_n(X)}{n!} t^n\,.
\end{equation}
In particular, the moments can be recovered from derivatives of $M_X(t)$ evaluated at $t=0$,
\begin{equation}
    m_n(X) = \left.\frac{d^n}{dt^n} M_X(t)\right|_{t=0}.
\end{equation}

Alternatively, one can characterize a random variable $X$ in terms of its cumulants $c_n$, defined through the cumulant-generating function
\begin{equation}
    K_X(t) = \sum_{n=1}^{\infty} c_n \frac{t^n}{n!},
\end{equation}
which is related to the moment-generating function by
\begin{equation}
    K_X(t) = \log M_X(t) = \log \mathbb{E}[e^{tX}].
\end{equation}
Using that $M_X(t)=e^{K_X(t)}$, we can find a relation between moments and cumulants as follows:
\begin{equation}
    m_n(X)= \left.\frac{d^n}{dt^n} e^{K_X(t)}\right|_{t=0}.
\end{equation}
This gives, for the first few moments,
\begin{align} \label{eq:MomentsFromCumulants}
    m_1(X) &= c_1(X), \nonumber\\
    m_2(X) &= c_2(X) + c_1^2(X), \nonumber\\
    m_3(X) &= c_3(X) + 3c_2(X) c_1(X) + c_1^3(X), \nonumber\\
    m_4(X) &= c_4(X) + 4c_3(X) c_1(X) + 3c_2^2(X) + 6c_2(X) c_1^2(X) + c_1^4(X).
\end{align}
Solving for the cumulants, one finds
\begin{align}
    c_1(X) &= m_1(X), \nonumber\\
    c_2(X) &= m_2(X) - m_1^2(X), \nonumber\\
    c_3(X) &= m_3(X) - 3m_2(X)m_1(X) + 2m_1^3(X), \nonumber\\
    c_4(X) &= m_4(X) - 4m_3(X)m_1(X) - 3m_2^2(X)
    + 12m_2(X)m_1^2(X) - 6m_1^4(X).
\end{align}

The relations \eqref{eq:MomentsFromCumulants} can be written in a general form using the concept of partitions. 
A partition $\pi$ of the set $\{1,\dots,n\}$ is a decomposition into disjoint subsets (called blocks) whose union is $\{1,\dots,n\}$. 
We denote by $|\pi|$ the number of blocks of the partition, and by $\mathcal{P}(n)$ the set of all partitions of $\{1,\dots,n\}$. 

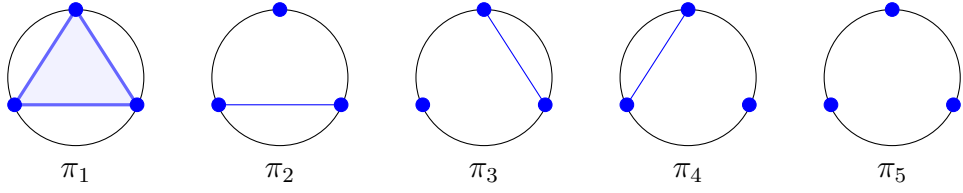
\begin{figure}[h!]
    \centering
    \begin{subfigure}[b]{\textwidth}
    \centering
    \begin{tikzpicture}[scale=1.8]
        \draw (-0.5,0) circle(0.5);
        \filldraw[color=blue!60, fill=blue!5, very thick](-0.95,-0.2) -- (-0.05,-0.2) -- (-0.5,0.5) -- cycle;
        \draw[fill=blue,blue](-0.95,-0.2) circle [radius=0.05];
        \draw[fill=blue,blue](-0.05,-0.2) circle [radius=0.05];
        \draw[fill=blue,blue](-0.5,0.5) circle [radius=0.05];
        \node at (-0.5,-.7){$\pi_1$};
        %\node at (0.25,0){$+$}; 
        
        \draw (1,0) circle(0.5);
        \draw[fill=blue,blue](1.45,-0.2) circle [radius=0.05];
        \draw[fill=blue,blue](0.55,-0.2) circle [radius=0.05];
        \draw[fill=blue,blue](1,0.5) circle [radius=0.05];
        \draw[blue] (1.45,-0.2)--(0.55,-0.2);
        \node at (1,-.7){$\pi_2$};

        \draw (2.5,0) circle(0.5);
        \draw[fill=blue,blue](2.95,-0.2) circle [radius=0.05];
        \draw[fill=blue,blue](2.05,-0.2) circle [radius=0.05];
        \draw[fill=blue,blue](2.5,0.5) circle [radius=0.05];
        \draw[blue] (2.95,-0.2)--(2.5,0.5);
        \node at (2.5,-.7){$\pi_3$};

        \draw (4,0) circle(0.5);
        \draw[fill=blue,blue](4.45,-0.2) circle [radius=0.05];
        \draw[fill=blue,blue](3.55,-0.2) circle [radius=0.05];
        \draw[fill=blue,blue](4,0.5) circle [radius=0.05];
        \draw[blue] (3.55,-0.2)--(4,0.5);
        \node at (4.0,-.7){$\pi_4$};

        \draw (5.5,0) circle(0.5);
        \draw[fill=blue,blue](5.95,-0.2) circle [radius=0.05];
        \draw[fill=blue,blue](5.05,-0.2) circle [radius=0.05];
        \draw[fill=blue,blue](5.5,0.5) circle [radius=0.05];
        \node at (5.5,-.7){$\pi_5$};
    \end{tikzpicture}
    
    \label{fig: partition_diag_n3}
    \end{subfigure}
    
    \captionsetup{justification=raggedright,singlelinecheck=false}
    \caption{Diagrammatic representation of the 5 partitions of the set $\{1,2,3\}$ listed in \eqref{eq:partitions}. The elements are placed on a circle in cyclic order, and blocks of a partition are represented by chords connecting the corresponding points. }
    \label{fig:partition_diag1}
\end{figure}

For example, the partitions of the set $\{1,2,3\}$ are
\begin{equation} \label{eq:partitions}
\begin{aligned}
 \pi_1 &= \{\{1,2,3\}\}, \\
 \pi_2 &= \{\{1\},\{2,3\}\}, \\
 \pi_3 &= \{\{2\},\{1,3\}\}, \\
 \pi_4 &= \{\{3\},\{1,2\}\}, \\
 \pi_5 &= \{\{1\},\{2\},\{3\}\}.
\end{aligned}
\end{equation}
Each partition corresponds to a way of grouping the indices into blocks. 
They can be represented graphically as indicated in Fig.~\ref{fig:partition_diag1}.

The relation between moments and cumulants can then be expressed as a sum over partitions:
\begin{equation}
    m_n(X) = \sum_{\pi \in \mathcal{P}(n)} \prod_{B \in \pi} c_{|B|}(X),
\end{equation}
where the product runs over the blocks $B$ of the partition $\pi$, and $|B|$ denotes the size of each block.

Applying the general formula to $n=3$, we obtain:
\begin{align}
m_3(X)
&= \sum_{\pi \in \mathcal{P}(3)} \prod_{B \in \pi} c_{|B|}(X) \nonumber\\
&= \prod_{B\in\pi_1} c_{|B|}(X)
 + \prod_{B\in\pi_2} c_{|B|}(X)
 + \prod_{B\in\pi_3} c_{|B|}(X)
 + \prod_{B\in\pi_4} c_{|B|}(X)
 + \prod_{B\in\pi_5} c_{|B|}(X) \nonumber\\
&= c_3(X)
  + c_1(X)c_2(X)
  + c_1(X)c_2(X)
  + c_1(X)c_2(X)
  + c_1(X)^3 \nonumber\\
&= c_3(X) + 3\,c_1(X)c_2(X) + c_1(X)^3\,.
\end{align}

The expansion of moments in terms of cumulants can also be generalized to moments involving several different random variables:
\begin{equation} \label{eq:momentFromCumulants}
m_n(X_1,X_2, \cdots X_n)=\mathbb{E}[X_1 X_2 \cdots X_n]
=
\sum_{\pi \in \mathcal{P}(n)}
\prod_{B \in \pi}
c_{|B|}\big(X_{B(1)},X_{B(2)},\dots,X_{B(|B|)}\big),
\end{equation}
where $\mathcal{P}(n)$ denotes the set of all partitions of $\{1,\dots,n\}$, and
$\{B(1),B(2),\dots,B(|B|)\}$ are the elements of a block $B$ of size $|B|$. 

For the first few orders, this formula gives
\begin{align}
\mathbb{E}[X_1]
&= c_1(X_1), \nonumber\\
\mathbb{E}[X_1X_2]
&= c_2(X_1,X_2)+c_1(X_1)c_1(X_2), \nonumber\\
\mathbb{E}[X_1X_2X_3]
&= c_3(X_1,X_2,X_3)
+c_1(X_1)c_2(X_2,X_3)
+c_1(X_2)c_2(X_1,X_3)
\nonumber\\
&\quad
+c_1(X_3)c_2(X_1,X_2)
+c_1(X_1)c_1(X_2)c_1(X_3).
\end{align}
Equivalently, the first cumulants can be written in terms of moments as
\begin{align}
c_1(X_1)
&= \mathbb{E}[X_1], \nonumber\\
c_2(X_1,X_2)
&= \mathbb{E}[X_1X_2]
-\mathbb{E}[X_1]\mathbb{E}[X_2], \nonumber\\
c_3(X_1,X_2,X_3)
&= \mathbb{E}[X_1X_2X_3]
-\mathbb{E}[X_1]\mathbb{E}[X_2X_3]
-\mathbb{E}[X_2]\mathbb{E}[X_1X_3]
\nonumber\\
&\quad
-\mathbb{E}[X_3]\mathbb{E}[X_1X_2]
+2\,\mathbb{E}[X_1]\mathbb{E}[X_2]\mathbb{E}[X_3].
\end{align}

\paragraph{Definition 4: Independence between random variables.}
Given a probability space $(\Omega,\Sigma,\mu)$, and two random variables
$X:\Omega \rightarrow \mathbb{R}$ and $Y:\Omega \rightarrow \mathbb{R}$,
$X$ and $Y$ are said to be independent if the events they generate are independent. More precisely, for any measurable sets $I_A,I_B \subset \mathbb{R}$,
\begin{equation}
    \mu\!\left(X^{-1}(I_A)\cap Y^{-1}(I_B)\right)
    =
    \mu\!\left(X^{-1}(I_A)\right)
    \mu\!\left(Y^{-1}(I_B)\right).
\end{equation}

Equivalently, independence implies the factorization of mixed moments:
\begin{equation}
    \mathbb{E}[X^n Y^m]
    =
    \mathbb{E}[X^n]\,
    \mathbb{E}[Y^m],
\end{equation}
for all non-negative integers $n,m$.

An alternative and particularly useful characterization of independence is formulated in terms of cumulants. A fundamental property of classical cumulants is that all mixed cumulants involving independent random variables vanish. In other words, if $X$ and $Y$ are independent, then
\begin{equation}
    c_n(X,\dots,X,Y,\dots,Y)=0,
\end{equation}
for any cumulant containing both $X$ and $Y$. 

\subsection{Algebraic formulation of classical probability spaces}

An alternative way to formulate probability theory is in terms of the algebra of random variables itself, rather than the underlying Kolmogorov triple $(\Omega,\Sigma,\mu)$. In the classical setting, random variables can be viewed as measurable functions
$ X:\Omega \rightarrow \mathbb{R}$ and the collection of bounded measurable functions forms a commutative algebra $\mathcal A$ under pointwise addition and multiplication.

The probability measure $\mu$ naturally defines a linear functional
\begin{equation}
    \varphi(X)=\mathbb{E}[X]
    =
    \int_\Omega X(\omega)\, d\mu(\omega),
\end{equation}
which plays the role of expectation value. One can therefore reformulate classical probability theory in terms of the pair $(\mathcal A,\varphi)$, abstracting away the underlying sample space and focusing directly on the algebraic relations between random variables and their expectation values~\cite{Tao2014AlgebraicProbability, novak2012lecturesfreeprobability}. This viewpoint is particularly useful because it admits a natural non-commutative generalization, in which observables are represented by non-commuting operators rather than ordinary functions. This leads to the framework of non-commutative and free probability theory, which we discuss in Sec.~\ref{sec:freeprobability}.

\subsection{Classical Dynamical Systems as an Example of a Probability Space}

To emphasize the relevance of probability theory in physics, we briefly discuss how statistical ensembles associated with classical dynamical systems naturally provide examples of probability spaces. In the context of ergodic theory \cite{Halmos1956ErgodicTheory}, one studies the statistical properties of trajectories generated by deterministic time evolution.

\paragraph{Definition 5: Classical dynamical system.}
A classical dynamical system is a probability space $(\Omega,\Sigma,\mu)$ equipped with a family of time-evolution maps $T_t:\Omega \to \Omega$ satisfying: (i) $T_0=\mathrm{id}$; (ii) $T_{t+s}=T_t\circ T_s$ for all $t,s$; and (iii) $\mu(T_t^{-1}(A))=\mu(A)$ for every measurable set $A\in\Sigma$.

The map $T_t$ describes the deterministic time evolution of points in phase space, while the probability measure $\mu$ characterizes a statistical ensemble of initial conditions. Measure preservation expresses the fact that the dynamics conserves the probability distribution under time evolution. In this setting, observables are represented by measurable functions
\begin{equation}
    f:\Omega\to\mathbb{R},
\end{equation}
whose time evolution is induced by the dynamical map according to
\begin{equation}
    f_t(\omega)=f(T_t(\omega)).
\end{equation}
The expectation value of an observable is then given by
\begin{equation}
    \mathbb{E}[f]
    =
    \int_\Omega f(\omega)\, d\mu(\omega).
\end{equation}

%%%%%%%%%%%%%%%%%%%%%%%%%%%%%%%%%%%%%%%%%%%%%%%%%%
\bibliographystyle{JHEP}
\bibliography{ref.bib}

\end{document}